\documentclass[11pt,a4paper]{article}
\pdfoutput=1

\usepackage{jheppub}

\usepackage{amsmath,hyperref,amssymb}
\usepackage{graphicx,url,color}
\renewcommand{\b}{\bar}
\newcommand{\be}{{\bar{1}}}

\DeclareSymbolFont{AMSa}{U}{msa}{m}{n}
\DeclareSymbolFont{AMSb}{U}{msb}{m}{n}
\DeclareMathSymbol{\fieldR}{\mathalpha}{AMSb}{"52}

\newcommand\nn{{\nonumber}}
\newcommand{\beq}{\begin{equation}}
\newcommand{\eq}{\end{equation}}

\def\be{\begin{equation}}
\def\ee{\end{equation}}
\def\bea{\begin{eqnarray}}

\def\eea{\end{eqnarray}}

\def\lab{\label}

\def\o{\omega}
\def\le{\left}
\def\ri{\right}

\def\m{\mu}
\def\n{\nu}

\def\6{\partial}

\def\a{\alpha}
\def\b{\beta}
\def\lab{\label}

\def\eps{\epsilon}

\newcommand{\mc}{\mathcal}

\def\be{\begin{equation}}
\def\ee{\end{equation}}
\def\bea{\begin{eqnarray}}
\def\eea{\end{eqnarray}}
\def\lab{\label}
\def\nn{\nonumber}
\def\le{\left}
\def\ri{\right}
\def\6{\partial}
\def\eps{\epsilon}

\title{A magnetically induced quantum critical point in holography}
\author[a]{A. Gnecchi,}
\author[b]{U. Gursoy,}
\author[b]{O. Papadoulaki,}
\author[c]{C. Toldo,}
\affiliation[a]{Institute for Theoretical Physics, KU Leuven, \\ 3001 Leuven, Belgium}
\affiliation[b]{Institute for Theoretical Physics and Center for Extreme Matter and Emergent Phenomena,\\
Utrecht University, Buys Ballot Building (BBG), Princetonplein 5, 3584 CC Utrecht, The Netherlands}
\affiliation[c]{Department of Physics, Columbia University, 538 West 120th Street,\\
New York, NY 10027, USA}
\emailAdd{a.gnecchi@kuleuven.be}
\emailAdd{u.gursoy@uu.nl}
\emailAdd{o.papadoulaki@uu.nl}
\emailAdd{ct2673@columbia.edu}
\abstract{We investigate quantum critical points in a 2+1 dimensional gauge theory at finite chemical potential $\chi$ and magnetic field $B$. The gravity dual is based on 4D $\mathcal{N}=2$ Fayet-Iliopoulos gauged  supergravity and the solutions we consider---that are constructed analytically---are extremal, dyonic, asymptotically $AdS_4$ black-branes with a nontrivial radial profile for the scalar field. We discover a line of second order fixed points at $B=B_c(\chi)$ between the dyonic black brane and an extremal ``thermal gas'' solution with a singularity of good-type, according to the acceptability criteria of Gubser \cite{Gubser:2000nd}. The dual field theory is a strongly coupled nonconformal field theory at finite charge and magnetic field, related to the ABJM  theory \cite{Aharony:2008ug} deformed by a triple trace operator $\Phi^3$. This line of fixed points might be useful in studying the various strongly interacting quantum critical phenomena such as the ones proposed to underlie the cuprate superconductors. We also find curious similarities between the behaviour of the VeV  $\langle \Phi \rangle$ under B and that of the quark condensate in 2+1 dimensional NJL models.}
\arxivnumber{1604.04221}

\begin{document}
\maketitle

\section{Introduction and summary}
\lab{Intro}

Quantum criticality is proposed to play a fundamental role in solution to important open problems in physics, such as the high $T_c$ superconductivity \cite{Sachdev}. Strongly interacting fixed points can be obtained by tuning a certain coupling in these systems  such as the pressure, the doping fraction or the magnetic field to a critical value, see e.g. \cite{Sachdev:2011wg} for multiple examples. The characteristic energy scale $\Delta E$ that governs the spectrum of fluctuations in these systems vanishes as one approaches this critical point. If the critical point corresponds to  second or higher order then this results in a conformal field theory as an effective theory governing the dynamics around criticality.  
Quantum phase transitions essentially happen in two different ways. It can correspond to a level crossing or a limiting case of an avoided level crossing. The second case appears to be more common in the condensed matter systems \cite{Sachdev:2011wg}. 

On the other hand, the AdS/CFT correspondence \cite{Maldacena:1997re,Gubser:1998bc,Witten:1998qj} has proved, over the last two decades, to be one of the most effective methods in addressing the strongly interacting critical phenomena. In this paper, we take this route to analyse  quantum phase transitions at strong coupling, from a dual holographic view. As an example of a strongly interacting field theory one may consider the ABJM model \cite{Aharony:2008ug}, deformed by a bosonic, gauge invariant triple trace operator $\Phi^3$, and placed at finite charge $q$ and magnetic field $B$. The theory we consider in this paper is related to this model. In fact, we define the precise theory through  the dual gravitational dual background  \cite{Duff:1999gh,Cvetic:1999xp} that are analytic solutions to $\mathcal{N}=2$  $U(1)$-gauged (Fayet--Iliopoulos) supergravity in 4 dimensions \cite{Andrianopoli:1996cm}. The aforementioned triple trace deformation corresponding to a scalar field $\varphi(r)$ with a particular profile in the holographic coordinate $r$, that is determined by an integration constant $b$, which can be thought of as the value of the VeV of the corresponding bosonic gauge invariant operator $\Phi$. Thus, the solutions we consider in this paper are governed by three parameters: the VeV $b$, the charge $q$ (alternatively, the chemical potential $\chi$), and the magnetic field $B$. Had this solution to  $\mathcal{N}=2$  $U(1)$-gauged (Fayet--Iliopoulos) supergravity in 4 dimensions been completely equivalent to the corresponding M2 brane solution in 11D, would one confidently identify the field theory with the deformed ABJM model mentioned in the beginning of this paragraph. However, instabilities may arise for the scalar fields which are left outside of the truncation to 4D \cite{Donos:2011qt,Donos:2011pn,Almuhairi:2011ws}. Therefore, in the most general case the field theories dual to our solutions are strongly coupled, non conformal theories placed at finite charge $q$ and magnetic field $B$ that can be obtained from the deformed ABJM model by following the RG flow initiated by scalar VeVs corresponding to such instabilities.  

Four dimensional $\mathcal{N}=2$  Fayet--Iliopoulos supergravity allows for the existence of black branes in asymptotically locally $AdS_4$ space, preserving 2 real supercharges (1/4-BPS states)\cite{Cacciatori:2009iz}. Their generalization to non-supersymmetric and finite temperature solutions were first constructed in \cite{Klemm:2012yg, Toldo:2012ec}. There has been a lot of progress, recently, on holography for BPS solutions in $AdS_4$ from gauged Supergravity, leading to the microstate counting of 1/4-BPS black holes entropy \cite{Benini1}, \cite{Hristov1,Benini2,Hosseini1}. In these examples there exist an $AdS_2$ factor in the near horizon region of the supersymmetric solution, corresponding to an IR fixed point to which the conformal UV theory flows, as a result of the topological twist induced at the $AdS_4$ boundary by the presence of magnetic fields. Another related line of investigations in the literature involve a holographic study of the ABJM type models deformed by dynamical flavors \cite{Jokela:2013qya,Bea:2016fcj}. The latter paper \cite{Bea:2016fcj} also reports similar quantum critical behaviour in the ABJM model deformed by dynamical flavor degrees of freedom. Finally, dilatonic, charged and dyonic black-branes have been investigated in the holographic context in a series of papers  by Goldstein et al \cite{Goldstein:2009cv, Goldstein:2010aw} and \cite{Amoretti:2016cad}. 

We focus on two different types of such solutions in this paper: the first one is an asymptotically AdS, extremal and dyonic\footnote{We consider a theory in which two abelian electric-magnetic gauge fields are present. Our main subject of investigation will be systems that are electrically charged with respect to the first gauge field and magnetically charged with respect to the second. Even though they are not dyonic under the same gauge field, we use a broader definition of dyonic system and we refer to them as possessing generic electric and magnetic charges.} black brane solution with a horizon at a finite locus $r=r_h$. We denote this solution with a subscript ``BB" below. The second type of solution is horizonless dyonic ``thermal gas" solution that can be obtained by sending the horizon $r_h$ to a singularity $r_s$. We denote this solution with a subscript ``TG" below. Generically it is insufficient to treat these latter type of singular solutions in the classical gravity approximation. However, as shown in  \cite{Gubser:2000nd}, if the singularity can always be cloaked by a horizon, the two-derivative gravity approximation is able to capture interesting IR physics in the dual CFT at vanishing string coupling $g_s$ (corresponding to large N in the dual gauge theory)\footnote{We elaborate on details of the criteria in section \ref{Good}.}. We find that this latter requirement results in the following non-trivial conditions: 
\be\lab{good} 
q_{TG} = 0, \qquad b_{TG} = \pm 2^{-\frac74} \sqrt{|B|}\, .
\ee      
Having imposed these conditions on the TG solution, we then seek for possible phase transitions between the BB and the TG branches by considering the difference of free energies between these branches $\Delta F = F_{BB} - F_{TG}$. 

We find that this difference indeed vanishes at the critical locus 
\be\lab{pt1} 
|B| = B_c(\chi) = \frac{4\sqrt{2}}{3} \chi^2\, .
\ee      
As one approaches this locus, the difference of free energies vanishes quadratically and the difference of magnetizations and the VeVs of the scalar operator vanish linearly,
\be\lab{pt2} 
\Delta F \approx \frac{3\sqrt{3}}{2} \frac{(B-B_c)^2}{\chi}\ ,\qquad \Delta M\approx  \frac{3\sqrt{3}}{2} \frac{B-B_c}{8\chi}\ , \qquad \Delta b  \approx \frac{3\sqrt{3}}{16}\frac{(B-B_c)}{\chi}\,  
\ee      
signaling a {\em second order} quantum critical point. In particular, as we show below, the two solutions become the same as $B$ approaches $B_c$. The {\em order parameter} of this critical behavior can then be identified as either the  magnetization or the VeV of the scalar operator. The magnetization behaves linearly in B in the BB phase and as $\sqrt{B}$ in the TG phase: 
\be\lab{mag} 
m_{BB} = \frac{3\sqrt{3}}{\sqrt{2}} \frac{B}{|\chi|}, \qquad \,  m_{TG} = 3\, 2^{-\frac54} \sqrt{|B|}{\textrm sgn}(B)\, ,
\ee      
exhibiting a discontinuity in the derivative with respect to both $B$ and $\chi$ at the critical point. Similar scaling arise when one considers the VeV $\langle \Phi \rangle$ in the BB and the TG phases, as we show in section \ref{Thermo}. At this point, one should emphasize that there is no independent source for the operator $\Phi$ in the dual theory. Therefore the VeV is completely set by the intensive variables $B$ and $\chi$ (chemical potential corresponding to electric charge $q$) at vanishing $T$. 

It is tempting to relate the phase coalescence we find here to a confinement-deconfinement type critical behaviour as usually is the case with the Hawking-Page type transitions between a black brane and a thermal gas geometry. However,  we show in section \ref{extended} that this expectation is false. In particular, we calculate the Polyakov loop holographically, and show that it is finite on both backgrounds. A computation of the quark anti-quark potential  supports this conclusion. Finally, we determined the holographic entanglement entropy between a region and its complement in the dual field theory and observed that the thermal gas also corresponds to a ``deconfined'' state in the corresponding field theory along with the black brane phase. As we argue in section \ref{NJL}, the critical point is more similar to  formation of quark condensate in the 3D Nambu-Jona-Lasinio models with magnetic field, rather than a confinement-deconfinement type transition. 

\begin{figure}[t!]
\begin{center}
\includegraphics[scale=0.7]{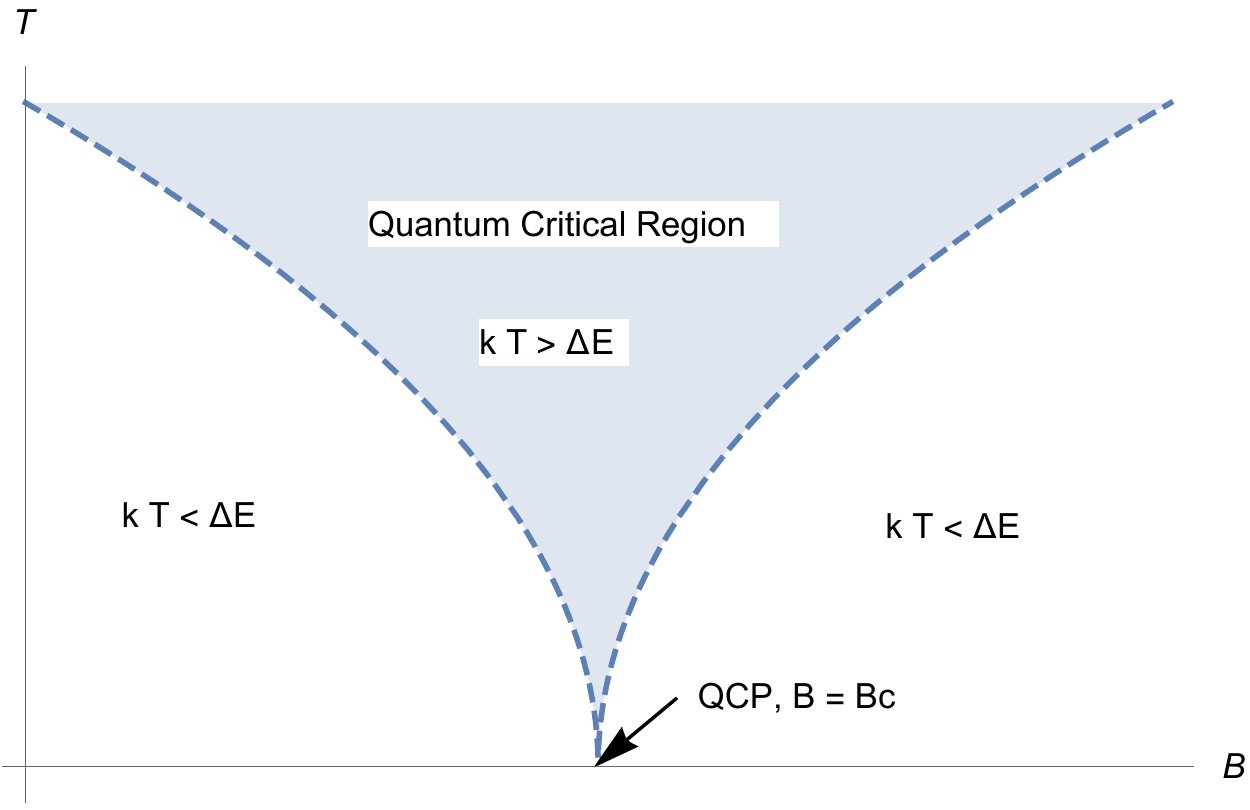}
 \end{center}
 \caption[]{The quantum critical region on a fixed $\chi$ slice. The boundary of the critical region, shown by the dashed line, is where the thermal fluctuations are of the same order as the intrinsic energy scale $\Delta E$. The latter vanishes as one approaches the critical point, signaling a quantum  phase transition.}
\label{fig1}
\end{figure}

We also consider the spectrum of fluctuations around these two type of solutions and find that the spectrum is gapped in the TG phase determined by a non-vanishing characteristic energy scale $\Delta E$ and that this energy scale vanishes in the BB phase. These findings are discussed in section \ref{flucs} and they are in accord with the expectations from quantum critical points mentioned above. In particular, the boundary of the {\em quantum critical region} in the phase space should be determined by the condition $\Delta T \sim T \sim \Delta E$ and $\Delta E$ should vanish as one approaches the critical point. This is because the  TG solution approaches the BB solution in the vicinity of the critical point and the latter has a zero mode. Just by dimensional analysis one can determine the boundary of the quantum critical region on a fixed $\chi$ slice of the phase diagram as $T \propto (B-B_c)^p/B_c^{p-1/2}$ where $p$ is a positive real number which we do not determine in this paper. The expected phase diagram on fixed $\chi$ slice is shown in figure \ref{fig1}.

The rest of the paper is organized as follows. In the next section we describe the gravity setting and introduce the dyonic black brane background. In particular section \ref{Good} discusses the singular limit of these black branes and outlines construction of the thermal gas backgrounds with a ``good" type singularity. We derive the good singularity condition (\ref{good}) in this section. In section \ref{Thermo}, we study the thermodynamics of the system in the mixed ensemble defined by finite electric chemical potential $\chi$ and magnetic field $B$ at vanishing temperature and establish the presence of the quantum critical point. At the end of this section, in subsection \ref{NJL}, we compare our findings with similar phenomena observed in 2+1 dimensional Nambu-Jona-Lasinio models. In particular we discuss qualitative similarities and dissimilarities in the profile of the condensate between our holographic model and the NJL models. In section \ref{flucs} we consider fluctuations around our backgrounds obtained by exciting point-like fields and extended  objects such as a Nambu-Goto string and minimal surfaces. Here we show that the quantum criticality we find is not associated with a confinement-deconfenement type. Finally in section  \ref{Discussion} we discuss the various implications of our findings in regard to applications in particle physics and condensed matter. We also give an outlook of the various routes one can extend our investigations. Several appendices detail our calculations.

\section{Gravity Set Up}
\lab{Gravity}

Our starting point is the Einstein-Maxwell-scalar theory with two gauge fields and one real scalar field, 
 ($\kappa^2=8\pi G_N$)
    \begin{eqnarray}\label{action}
     I=\frac{1}{\kappa^2}\int \sqrt{-g}&d^4x&\left(
    \frac{R}{2}- \frac1{2} \partial_{\mu}\varphi \partial^{\mu}\varphi
     -e^{\sqrt6 \varphi}\xi^3F^0_{\mu\nu}F^{0\,\mu\nu}-\frac{3}{\xi}e^{-\sqrt{2/3}\varphi}F^1_{\mu\nu}F^{1\,\mu\nu}+\right.\nn\\
     &&  -V_g(\varphi)\Big)+S_{GH}\ .
    \end{eqnarray}
This action is identical to the bosonic action of an $\mathcal{N} =2$ sector of $\mathcal{N} =8$ gauged supergravity, obtained by truncating the $SO(8)$ gauging to the $U(1)^4$ Cartan subgroup and further restricting to the diagonal $U(1)$ \cite{Duff:1999gh}. In the language of $\mathcal{N}=2$ gauged supergravity it corresponds to a Fayet-Iliopoulos, or R-symmetry gauging where the $U(1)_R\in SU(2)_R$ symmetry is made local. Moreover, the field content can be seen as a no-axions truncation of the $\mathcal{N} =2$ Supergravity special geometry described by the prepotential $F=-2i\sqrt{X^0 (X^1)^3}$ \cite{Cvetic:1999xp}, with the identification $z=X^0/X^1=e^{\sqrt{3/8}\varphi}$. 
The Gibbons- Hawking term is
\begin{equation}
S_{GH}=-\frac{1}{\kappa^2}\int d^3x \sqrt{-h} \Theta\ ,
\end{equation}
where $h_{mn}$ is the induced metric at the boundary and $\Theta$ is the trace of the extrinsic curvature of the boundary given by $\Theta _{\mu\nu}=-\frac{1}{2}({\nabla} _{\mu} n_{\nu}+{\nabla} _{\nu }n_{\mu})$; $n^{\mu}$ is the unit normal vector at the boundary pointing outwards.
The theory is specified by two constants $\xi_0,\xi_1$, a coupling $g$, and a scalar potential
\begin{eqnarray}
V_g(\varphi)&=&-\frac{3}{\ell_{AdS}^2}\cosh\left( \sqrt{\frac{2}{3}}\varphi(r)\right)
\ , \qquad \ell_{AdS}^2=\frac{3\sqrt3}{2g^2\sqrt{\xi_0(\xi_1)^3}}\ ,\qquad \xi=\sqrt{3\xi_0/\xi_1}\ ,\qquad
\end{eqnarray}
where $\ell_{AdS}$ is the AdS length scale. The theory admits a (supersymmetric) $AdS_4$ vacuum at the $\varphi=0$ locus. At this extremum the scalar field has mass $m^2_{\varphi}\ell_{AdS}^2=-2$, satisfying the Breitenlohner-Freedman bound. In particular, the mass of the scalar fits in the window $-\frac94<m^2_{\varphi}\ell_{AdS}^2<-9/4+1$ that allows for mixed boundary conditions for the scalar field at the boundary \cite{Papadimitriou:2007sj, Gnecchi:2014cqa}. From this point on we set $\xi_0=1/\sqrt2,\xi_1=3/\sqrt2$, $g=1$ thus $\xi=1,\ \ell_{AdS}=1$. In particular the radial direction $r$ will be considered as dimensionless below. One can easily recover the dimension of a given object by inserting appropriate powers of $\ell_{AdS}$ if needed. 

\subsection{Black branes\label{Branes solutions}}

We consider static, spherically symmetric black brane solutions of \eqref{action} supported by two {\em magnetic} gauge fields 
\begin{eqnarray}\lab{orsol}
A^\Lambda=\frac14 p^\Lambda (xdy-ydx)\ ,\qquad \Lambda=0,1\ ,
\end{eqnarray}
where $x$ and $y$ are the two spatial directions.

For reasons explained in section \ref{Thermo} we are rather interested in {\em dyonic} solutions with one electric and one magnetic charge, obtained by an electric-magnetic duality transformation only on the gauge field $A^0$. After the transformation the gauge fields are given as,  
\begin{eqnarray}
&&\tilde{F}^0=  \frac{q }{2(r-3b)^2}dr\wedge dt\ ,
\qquad
F^1= \frac B2 dx\wedge dy\ . 
\end{eqnarray}
The charges of the dualized configuration are related to the original one (\ref{orsol}) by
\begin{eqnarray}\label{magtomix}
q(p_0)=-p^0\equiv q 
\qquad
B(p^1)= p^1\equiv B
\ .
\end{eqnarray}
The duality transformation leaves the metric invariant, hence the solution is of the form as in \cite{Klemm:2012yg, Toldo:2012ec}. We use the parameterization as in \cite{Gnecchi:2014cqa}. The metric is
\begin{eqnarray}
\lab{metric}
ds^2&=&-\frac{f(r)}{\sqrt{H_0(r)H_1^3(r)}}dt^2+\sqrt{H_0(r)H_1^3(r)}\left(\frac{dr^2}{f(r)} +r^2(dx^2+dy^2)\right)
\end{eqnarray}
with 
\begin{eqnarray}\label{warp-factors}
&&H_0(r) = 1-\frac{3b}{r}\ , \quad H_1(r) = 1+\frac{b}{r}\ , \quad f(r)=\frac{c_1}{r} + \frac{c_2}{r^2} + r^2 H_0(r) H_1(r)^3 \ ,
\end{eqnarray}
while the metric coefficients are related to the charges as
\be\lab{defc1c2} 
c_1 = \frac{(p^0)^2 - (p^1)^2 }{2b}\, , \qquad c_2 = \frac{(p^0)^2 +3(p^1)^2 }{2}\, .
\ee
The scalar field has a radial profile
\begin{eqnarray}
\lab{scalar}
e^{\sqrt{8/3}\varphi}=\frac{r+b}{r-3b} \ .
\end{eqnarray}
which, from its asymptotic expansion at the boundary $r\rightarrow \infty$
\begin{eqnarray}\label{scalarexpans}
\varphi =  \frac{\varphi_-}{r} + \frac{\varphi_+}{r^2} + O(1/r^3) \,,  
\end{eqnarray}
reveals that it satisfies mixed boundary conditions: $\varphi_+=\frac1{\sqrt6} \varphi_-^2$ .
From the holographic point of view this correspond to the insertion of a multi trace deformation in the field theory \cite{Witten:2001ua,Berkooz:2002ug,Hertog:2004ns,Papadimitriou:2007sj}, which, in this case, is given by a triple trace deformation\footnote{\label{bc-choice}One has the freedom to interpret this deformation either as having Neuman boundary conditions for the scalar that correspond to a relevant deformation with a nonzero source at the field theory side or as mixed boundary conditions for the scalar that correspond to a marginal deformation with zero source at the field theory side. Depending of the boundary conditions the renomalization prescription changes and in the two cases we have to add different finite counterterms. These counterterms, in the end change the field theory interpretation despite the fact that the gravitation solutions look the same e.g.\cite{Benini2}\cite{Benini1}. In case though that one wants to preserve supersymmetry at the dual field theory, one cannot have mixed boundary conditions, so supersymmetry resolves the issue of the interpretation \cite{Freedman:2013ryh,Bobev:2011rv}.}
$\lambda \, \Phi^3$, with $\Delta_{\Phi}=1$, $\lambda=\frac1{\sqrt6}$. The dual operator corresponds to $\Phi\sim Tr[Z_1^\dagger Z^1-W^\dagger W]$, obtained after the identification of the bi-fundamental matter of the boundary theory as $Z^2=W^1=W^2=W$ \cite{Freedman:2013ryh}, giving an $\mc N=2$ truncation of ABJM. Multi traces are products of single trace operators, normalized canonically \cite{Witten:2001ua} s.t. $\langle \mc O \rangle= \mc O(N^0)$ as $N\to \infty$. Due to large-$N$ factorization, there is no mixing at leading order between single and multi trace operators.

It is important to note that the dyonic solution above is a solution of \eqref{action} with the modified kinetic terms for the gauge fields\footnote{The potential and the Einstein-Hilbert term remain invariant.}
\begin{eqnarray}
\mc L^{Dual}_{FF}= -e^{-\sqrt6 \varphi}\tilde{F}^0_{\mu\nu}\tilde{F}^{0\,\mu\nu}-3 e^{-\sqrt{2/3}\varphi}F^1_{\mu\nu}F^{1\,\mu\nu}\ .
\end{eqnarray}
We note that, in the holographic dual field theory we interpret $q$ as the charge density and $B$ as the magnetic field. Reinserting the dimensions, the correct identification is given by 
\be\lab{iden} 
q_{ft} = q\, \ell_{AdS}, \qquad e\,B_{ft} = B/\ell_{AdS},
\ee 
where $e$ is the electric charge in the dual 2+1 dimensional field theory. Finally, we provide an expression for the chemical potential associated with the conserved electric charge $q$, see  Appendix \ref{appA1} for details. The chemical potential is given by  
\begin{equation}\label{chi}
\chi =  - \int_{r_h}^{\infty} \tilde{F}^0_{tr} dr  = -\frac{q}{2(r_h-3b)}\, ,
\end{equation}
which we identify as the {\em electric chemical potential} after the aforementioned duality  transformation.  

\subsection{Good singularities and the thermal gas solution}
\lab{Good}

In addition to the black brane solution we described in the previous section, the action (\ref{action}) supports thermal gas type solutions \cite{Gursoy:2007er}, that are horizonless solutions with vanishing entropy. These solutions can be obtained from the black brane by sending the horizon location to the singularity, that are located at the zeros of the functions $H_0(r)$ and $H_1(r)$ in \ref{warp-factors}: 
\bea\lab{sing1} 
r_{s} &=& 3 b \qquad {\textrm for}\qquad b>0\, ,\qquad\qquad
\lab{sing2} 
r_{s} =- b \qquad {\textrm for} \qquad b<0\, .
\eea 
These solutions have curvature singularities that are expected to be resolved in the embedding to the full string theory. These solutions may still be acceptable  in the 3+1 dimensional supergravity reduction that we work with in this paper. In the AdS/CFT context, these solutions are dual to a well-defined state in the dual field theory if these singularities satisfy the Gubser's criteria \cite{Gubser:2000nd}. Indeed, such singular solutions which satisfy the criteria of \cite{Gubser:2000nd} are shown to correspond to the confined phase in the dual QCD-like gauge theories in \cite{Gursoy:2007cb,Gursoy:2007er}.

  \begin{figure}[t!]
 \begin{center}
\includegraphics[scale=0.48]{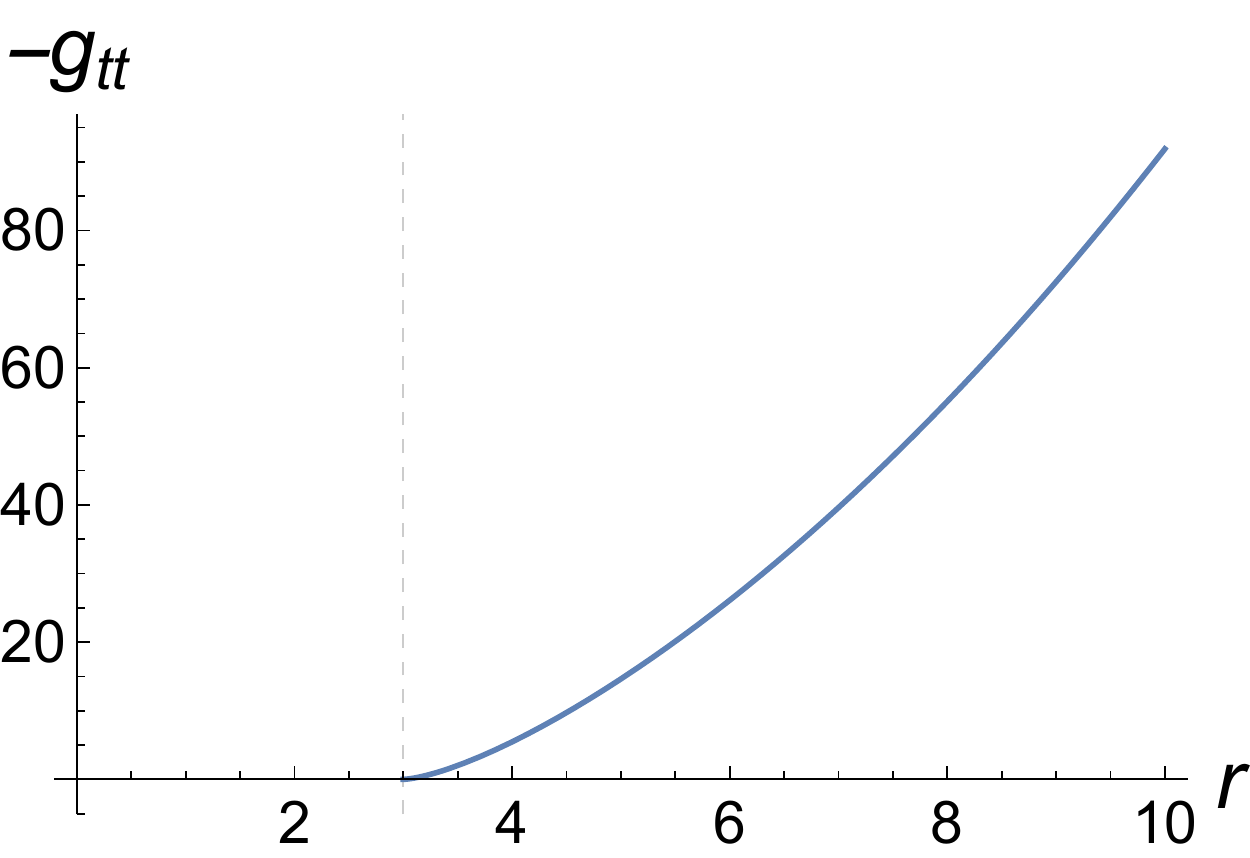} 
\includegraphics[scale=0.49]{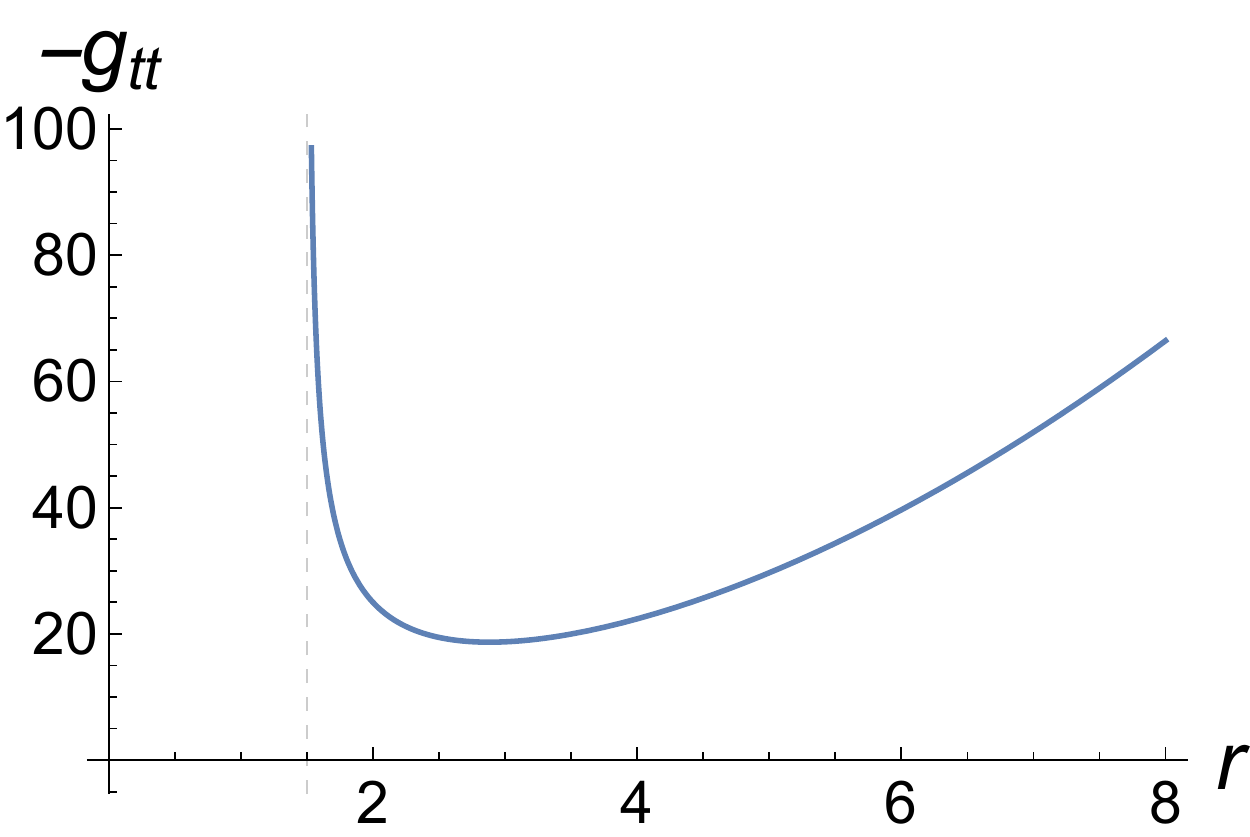}
 \end{center}
\caption[]{Example of warp factor $-g_{tt}$ for a good singularity (left: $B=8 \sqrt2$, $\chi= 1$, $r_h=3b=3$) and a bad one (right: $B=1$, $q= 5$, $b=0.5$). The dashed line indicates the location of the singularity. This coincides with the horizon, $r_h=r_{sing}$, in the plot to the left (good singularity); in this case there is no region where $-g_{tt}'(r)<0$. In the plot on the right (bad singularity), on the contrary, there exists a region where $-g_{tt}'(r)<0$ (for instance $r<2$).  In both cases of good and bad singularities, the warp factors $g_{yy}$ and $g_{xx}$ go to zero at $r=r_{sing}$ and the curvature invariants such as the Kretschmann scalar $R_{\mu \nu \rho \sigma}R^{\mu \nu \rho \sigma}$ diverge at $r=r_{sing}$. }
\label{figpotential}
\end{figure}

\begin{figure}[t!]
 \begin{center}
\includegraphics[scale=1]{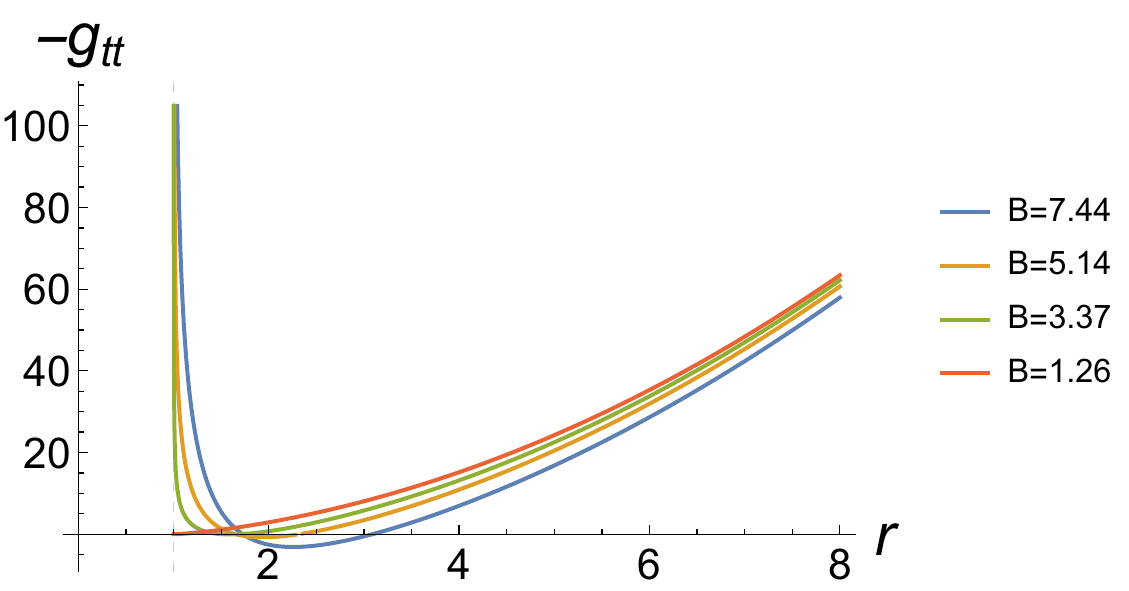} 
 \end{center}
\caption[]{Example of warp factor $-g_{tt}$ for black branes (blue, orange and green lines) and good singularity (red line). The curves corresponds to $b=0.33$ and $\chi=1$ and are drawn for the following values of magnetic charge: $B=7.44, 5.14, 3.37, 1.26$. In all cases the singularity is located at $r=r_s=1$; here the curvature invariants such as the Kretschmann scalar $R_{\mu \nu \rho \sigma}R^{\mu \nu \rho \sigma}$ diverge. The plot shows the family of black branes approaching the good singularity solution (thermal gas) as $B$ approaches the value given by \eqref{bsoltg}, at fixed $b$ and $\chi$. In the limit, the horizon is pushed to the singularity $r_s= 3b=r_h$.}
\label{figpotential2}
\end{figure}

The Gubser criterion requires the singular solution be obtainable from a black-hole in the limit that the horizon approaches the singularity $r_h\to r_s$. To be specific, we consider $b>0$. 
The criterion can be expressed in different ways depending on which parameters in the solution one keeps constant, that will eventually correspond to the choice of the thermodynamic ensemble. As will become clear in the next section, we find appropriate to work with the mixed ensemble where we keep constant the magnetic charge $B$ and the electric chemical potential $\chi$, defined in (\ref{chi}). Thus, the black brane solution is specified in terms of  $(b,B,q)$. Consider the family of black branes with horizon $r_h=3b+\epsilon$, for small $\epsilon$. For $b>0$, the horizon $r_h$ satisfies $f(r_h)=0$ where $f$ is defined in \eqref{warp-factors}. 
This equation, with $r_h=3b+\epsilon$, can be analytically solved in $B$. For small epsilon $\epsilon\ll1$, one finds 
\begin{eqnarray}\label{TG-limit}
|B|= 8\sqrt2 b^2 +\frac{(6b^2+\chi^2)\epsilon}{\sqrt2 b }+{\cal O}(\epsilon^2)\ .
\end{eqnarray}
This is an analytic expression valid for $\epsilon>0$, which determines the horizon of the black brane. Under the crucial assumption that the magnetization $\chi$ remains finite (see Sec. \ref{ThermoTG}), we analytically continue eq. $f(r_h)=0$ to $\epsilon=0$ and we take this limit as the defining relation of the \lq\lq{}good singularity\rq\rq{}. 
Namely, we define the thermal gas as the subset in parameter space $(B,b_{TG},\chi)$ defined by  
\be\lab{bsoltg} 
b_{TG} =  2^{-\frac74}\sqrt{|B|}\, . 
\ee
Using (\ref{chi}), we find that the charge of the thermal gas solution $q_{TG}$  vanishes linearly in the limit $\eps\to 0$: 
\be\lab{qsoltg} 
q_{TG} = - 2\chi\, \eps \to 0\, .
\ee
Notice that for the thermal gas the dependence on $\chi$ drops from the metric:
\begin{eqnarray}\lab{mettg}
ds^2_{TG}&=&-e^{-\sqrt6\varphi}(r^2+6b r+21 b^2)dt^2+e^{\sqrt{6}\varphi}dr^2(r^2+6b r+21 b^2)^{-1}+\nn\\
&&e^{\sqrt{6}\varphi}(r-3b)^2(dx^2+dy^2)\ ,\nn\\
e^{\sqrt{6}\varphi}&=&\left(\frac{r+b}{r-3b}\right)^{3/2}\ .
\end{eqnarray}
Since $B,b$ are related, a suitable set of parameters for the thermal gas is $(B,\chi,T)$, where $T$ is the temperature of the gas. Together with $\chi$, they are moduli of the thermal gas solution. Notice it has the same number of parameters as the black brane non-singular solution. 

Before concluding this section, let us mention the fact that the definition for a good singularity requires that the gravitational force acting on an uncharged probe particle is always attractive. More precisely, the radial motion of an uncharged particle with zero angular momentum and energy $E$ is determined by the equation (see for example \cite{Carroll:1997ar})
\begin{equation}
\frac12 \left( \frac{dr}{d\tau} \right)^2-V_{eff}(r) = \frac12 E^2\,,
\end{equation}
where the effective potential is given by
\begin{equation}
V_{eff}(r)= -\frac12 g_{tt}\,.
\end{equation}
Hence the requirement that the force on a probe particle is never repulsive translates in $ -dg_{tt} /dr >0$ throughout the spacetime. We illustrate this in Fig. \ref{figpotential}. In Fig \ref{figpotential2}, instead, we show how a good singularity (red line) can be obtained as a limit of regular black brane configurations, by tuning the value of the magnetic field $B$ (blue, yellow and green lines are non extremal black branes with two regular horizons). 

\section{Thermodynamic quantities and the quantum critical point}
\lab{Thermo}

\subsection{Thermodynamics of the black brane} 

Thermodynamic properties of gravity solutions in asymptotic Anti de Sitter space can be obtained via holographic renormalization of the on-shell action and gravitational stress-energy tensor. We have derived these quantities in Appendix \ref{appA1}, to which we refer for the following relations and identities.

The free energy of any black brane solution of Sec. \ref{Branes solutions} is 
\be\lab{fbb1} 
F_{BB} = M_{BB} - T S_{BB} + q_{BB} \chi\, ,
\ee   
where $M_{BB}$ is the mass of the black brane 
\be\lab{mbb} 
M_{BB} =    \frac{B^2- q_{BB}^2}{4b_{BB}}\, ,
\ee 
$T\,,S_{BB}$ are its temperature and entropy, $q_{BB}$ is the electric charge carried by the black brane, $\chi$ is the chemical potential given in equation (\ref{chi}). We are interested in the thermodynamics of these solutions in the mixed ensemble, defined by the free energy relation
\be\lab{diffbb}
dF_{BB}  = -S_{BB} dT +  q_{BB} d \chi\, + m_{BB} dB\, ,
\ee
therefore the independent thermodynamic variables are $T$, $\chi$ and $B$. 

Furthermore, we will restrict to the case of vanishing temperature, $T=0$.  In order to eliminate $r_h$, $q$ and $b$ in the expressions we will make use of (\ref{chi}), the  horizon equation $f(r_h)= 0$ ($f$ is as in (\ref{warp-factors})) and the extremality condition $f'(r_h)=0$.   
We obtain two interesting ways to express the free energy. First, we can solve the aforementioned equations in terms of $B$ and $q$ and obtain: 
\bea 
\lab{Beq}
B^2 &=&  \frac12\le( -5b^4 - 12b^3r_h - 6b^2r_h^2+ 4b r_h^3 + 3r_h^4\ri) \, ,\\ 
   \lab{qeq}
 q^2 &=& \frac{27 b^4}{2} + 18 b^3r_h-3 b^2
   r_h^2 -6 b r_h^3+\frac{3 r_h^4}{2}\, .
   \eea
Substitution in (\ref{fbb1}) then gives 
\be\lab{fbb2} 
F_{BB} = \frac14\le(5r_h-7b\ri)\le(b+r_h\ri)^2\, .
\ee   
This form is useful to check the zeros of the free energy. The function in \eqref{fbb2} clearly has a quadratic zero at  $r_h = -b$, which corresponds to the singularity when $b<0$, and this gives a consistency check. When $b>0$, it appears there is another zero at $r_h = 7b/5$, however this value for $r_h$ is smaller than the actual singularity $r_s= 3b$, so in fact the free energy has only one zero that is given by $r_h = -b$.  It is also clear from the formula above that the free energy is positive definite. 

It is more appropriate, for thermodynamics studies, to express the free energy in terms of the correct variables of mixed ensemble, namely $(T,B,\chi)$. To do so we just need to solve the equations above, this time in terms of $\chi$ and $B$. We obtain the following expressions: 
\bea\lab{rsol} 
r_h &=& \frac{9B^2+160 \chi^4}{64 \sqrt{6} |\chi|^3} \, ,\\
\lab{qsol}
q_{BB} &=& \frac{-9B^2+32 \chi^4}{8 \sqrt{6} \chi^2} \, \textrm{sgn}(\chi) \, ,\\
\lab{bsol}
b_{BB} &=&  \sqrt{\frac32} \frac{32 \chi^4- 3B^2}{64 |\chi|^3} \,.
\eea  
These solutions are valid only for $\chi\neq 0$. The free energy of the black brane then follows as,  
\be\lab{fbb3} 
F_{BB} = \frac{27B^2+32 \chi^4}{24 \sqrt{6} |\chi|}  \, .
\ee   
This form also makes it obvious that the free energy of the black brane is positive definite. Now one can check the first law of thermodynamics (\ref{diffbb}). The charge density $q$ should be obtained as 
\be\lab{check1} 
q_{BB} =  \frac{\6 F_{BB}}{\6\chi} \bigg|_{B,T} .\, 
\ee  
This indeed matches (\ref{qsol}) perfectly, hence the first law is satisfied. This provides another non-trivial check on our calculations.  Finally, the magnetization of the black brane solution is obtained as 
\be\lab{msol} 
m_{BB} =  \frac{\6 F_{BB}}{\6B} \bigg|_{\chi,T}  = 3\sqrt{\frac32}\, \frac{B }{|\chi|} \, . 
\ee  
We find that the magnetization of the black brane grows linearly with B. 

\subsection{Thermodynamics of the thermal gas\label{ThermoTG}} 

As explained in section \ref{Good}, the thermal gas solution is obtained from the black brane by sending the horizon $r_h$ to one of the singularities. In the following, in particular, we take $b>0$\footnote{Good singularities with $b<0$ require $B=0$ instead, thus, in this ensemble (in which the magnetic field is fixed) they do not compete with regular black branes, the latter generically having $B\neq0$.}. 
Consider then the limit of a black brane with horizon $r_h = 3b +\eps$, as $\eps\rightarrow0$. In the grand canonical ensemble the states have fixed magnetic potential, thus $\chi_{TG} = \chi_{BB} = \chi$, which remains finite in the limit. As explained in section \ref{Good}, this requirement implies the following relation between the parameter $b_{TG}$ of the thermal gas and the magnetic field $B$
\be\lab{bsoltg2} 
b_{TG} = + 2^{-\frac74}\sqrt{|B|}\, , 
\ee
 and it also leads to a vanishing electric charge, $q_{TG}=0$, see equation (\ref{qsoltg}).  
The temperature of the thermal gas is a moduli parameter, which can be set to any positive value. To match the black brane solution above, then, we choose $T_{TG}=0$. 
The entropy for the thermal gas also vanishes in this limit as, 
\be\lab{Slim} 
S_{TG} = 16\pi 2^{-\frac{21}{8}} |B|^{\frac34} \sqrt{\eps} \to 0 \, . 
\ee
In the appendix we compute the thermal gas free energy by holographic renormalization. We have verified that the same result is obtained by substituting in equation \eqref{fbb1} the expressions \eqref{qsoltg} and \eqref{Slim}, and imposing the defining relation \eqref{bsoltg2}. We arrive at 
\be\lab{ftg} 
F_{TG}  = M_{TG}  = \frac{B^2}{4b_{TG}} = 2^{-\frac14} |B|^{\frac32}\, .
\ee 
This result is clearly consistent with the first law of thermodynamics: the charge of the thermal gas solution obtained by the variation with respect to $\chi$ trivially vanishes, just like (\ref{qsoltg}). Moreover, the magnetization is given by 
\be\lab{msoltg} 
m_{TG} =  \frac{\6 F_{BB}}{\6B} \bigg|_{\chi,T}  = 3\cdot 2^{-\frac54} \sqrt{|B|}\, \textrm{sgn}(B)\, . 
\ee  
We note that the qualitative difference between the black brane (\ref{msol}) and the thermal gas (\ref{msoltg}) magnetization: the former is linear in B whereas the latter grows like the square root of B. 

\subsection{Difference of free energies and the quantum critical point} 

In the work presented so far, we have introduced all relevant physical quantities needed to study the thermodynamics phase space. We will proceed now to investigate possible phase transitions between black brane and thermal gas solutions. 

In order to determine whether a phase transition occurs, we consider the difference of free energies (\ref{fbb3}) and (\ref{ftg}):
\be\lab{fdif} 
\Delta F = F_{BB} - F_{TG}  = \frac{27B^2+32 \chi^4}{24 \sqrt{6} |\chi|} - 2^{-\frac14} |B|^{\frac32}\, .
\ee
We note that it is even under $B\rightarrow -B$ and $\chi\rightarrow -\chi$ independently. This means, as it should be, that the free energy is C-even and P-even. We can consider analogous checks for the difference of magnetizations 
\be\lab{mtot} 
m = \frac{\6 \Delta F}{\6 B} \bigg|_{\chi,T}  = \frac{3 \sqrt{6}}{8} \, \frac{B }{|\chi|} -  3\, 2^{-\frac54} \sqrt{|B|} \, \textrm{sgn}(B)\, ,
\ee
which is C-even and P-odd, and the difference of electric charges 
\be\lab{qtot}
q =  \frac{\6 \Delta F}{\6\chi} \bigg|_{B,T}  = q_{BB} =  \frac{-9B^2+32 \chi^4}{8 \sqrt{6} \chi^2} \, \textrm{sgn}(\chi) \ ,\, 
\ee
yielding that the total charge is P-even and C-odd. 

Finally, we find that the free energies of the TG and the BB phases become equal at 
\be\lab{pt} 
|B_c| = \frac{4\sqrt{2}}{3} \chi^2\, ,
\ee
corresponding to the zero of (\ref{fdif}).
Therefore, for every value of $\chi$ (except $\chi=0$) we find a non-analytic behaviour in the free energy at a finite magnetic field given by (\ref{pt}). Quite interestingly, this non-analytic behaviour is of {\em second order}. In fact, by expanding the difference of free energy (\ref{fdif}) near the critical point, we obtain 
\be\lab{fdifx} 
\Delta F = 3\sqrt{\frac32} \frac{(B - B_c)^2}{|\chi|} + {\cal O}(B - B_c)^3 \, .
\ee
In order to exhibit the discontinuity in the free energy we can directly compare the free energies and their derivatives at the critical point. We find that, even though the free energies and their first derivatives are continuous between the two phases, the second derivative jumps by an excess amount $\frac{3}{8}\sqrt{3/2}/|\chi|$ from the TG phase to the BB phase at $B=B_c$. Our example corresponds to the limiting case of an avoided level crossing, in the sense that the TG phase always wins over the BB phase everywhere in the phase space except the critical point $B=B_c$ where their free energies become equal. This is still called a ``quantum phase transition" according to the definition utilised in \cite{Sachdev:2011wg}. 
However we avoid using the term ``transition'' in this paper, as it sounds more appropriate for an actual level crossing. 

The excess magnetization (\ref{mtot}) can be expanded near the critical point as, 
\be\lab{mdifx} 
\Delta m = 3\sqrt{\frac32} \frac{(B - B_c)}{8|\chi|} + {\cal O}(B - B_c)^2 \, ,
\ee
showing that the difference of magnetization between the two phases vanishes linearly at the critical point. We note that the excess charge also vanishes, although quadratically, precisely at the critical point: 
\be\lab{qdifx} 
\Delta q = -3\sqrt{\frac32} \frac{(B - B_c)^2}{\chi^2} \, \textrm{sgn}(\chi) + {\cal O}(B - B_c)^2 \, .
\ee 
These results are another non-trivial check of our previous calculations. It is indeed expected that, at a second order critical point,  two competing solutions become the same (as opposed to a first order point where two different, competing states coexist). The solutions we consider are completely specified by $S$, $m$ and $q$. The entropies $S_{BB}$ and $S_{TG}$ vanish at the critical point (to check that $S_{BB}$ vanish one has to impose on the black brane parameters \eqref{rsol}-\eqref{bsol} the criticality condition \eqref{pt}), hence they are the same. As we have seen in (\ref{mdifx}) the magnetizations also become the same. For consistency of a second order critical point then the charges should also become the same. Since the charge of thermal gas vanishes, the critical behaviour then should happen when the charge of the black brane also vanishes, as nicely confirmed above.   

The difference between the vacuum expectation values of the condensates in the thermal gas phase, $b_{TG}$ in (\ref{bsoltg2})  and the black brane phase, $b_{BB}$ in (\ref{bsol}) also vanishes linearly as 
\be\lab{bdif} 
\Delta b = b_{TG} - b_{BB} = \frac{3\sqrt{3}}{16}\frac{(B-B_c)}{\chi} + {\cal O}(B - B_c)^2
\, .
\ee 

 \begin{figure}[t!]
 \begin{center}
\includegraphics[scale=0.7]{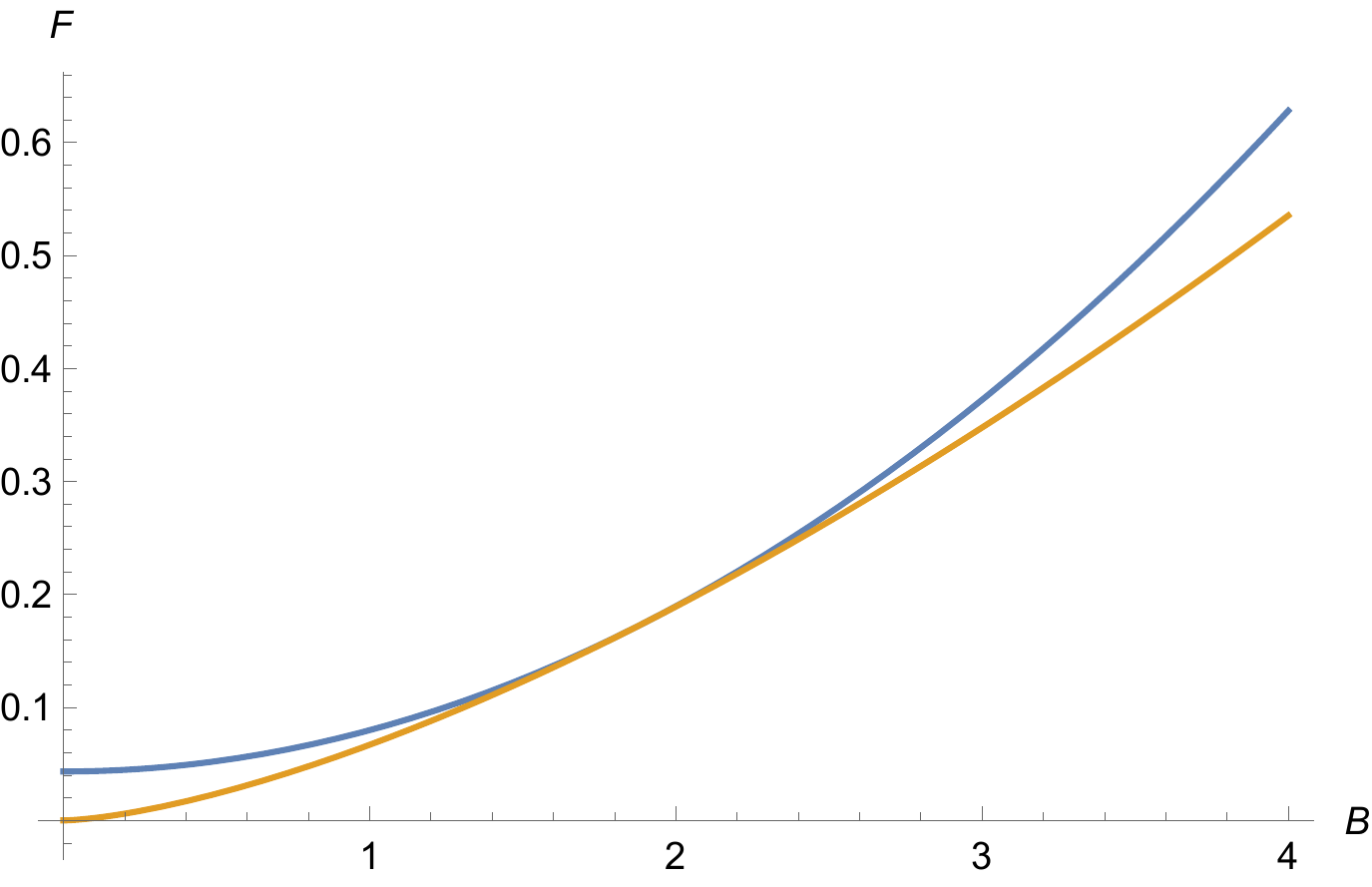}
 \end{center}
\caption[]{ Plot of the free energies of the black brane and the solitonic solution for $\chi =1$, as functions of $B$, the blue line is the black brane and the yellow line is the thermal gas.  The thermal gas is thermodynamically favoured everywhere in the $(B,\chi)$ region. At the critical line, defined as $B=\frac{4\sqrt2}{3}\chi^2$, black brane and thermal gas coincide.} 
\label{fig2}
\end{figure}

\begin{figure}[t!]
\begin{center}
\includegraphics[scale=0.7]{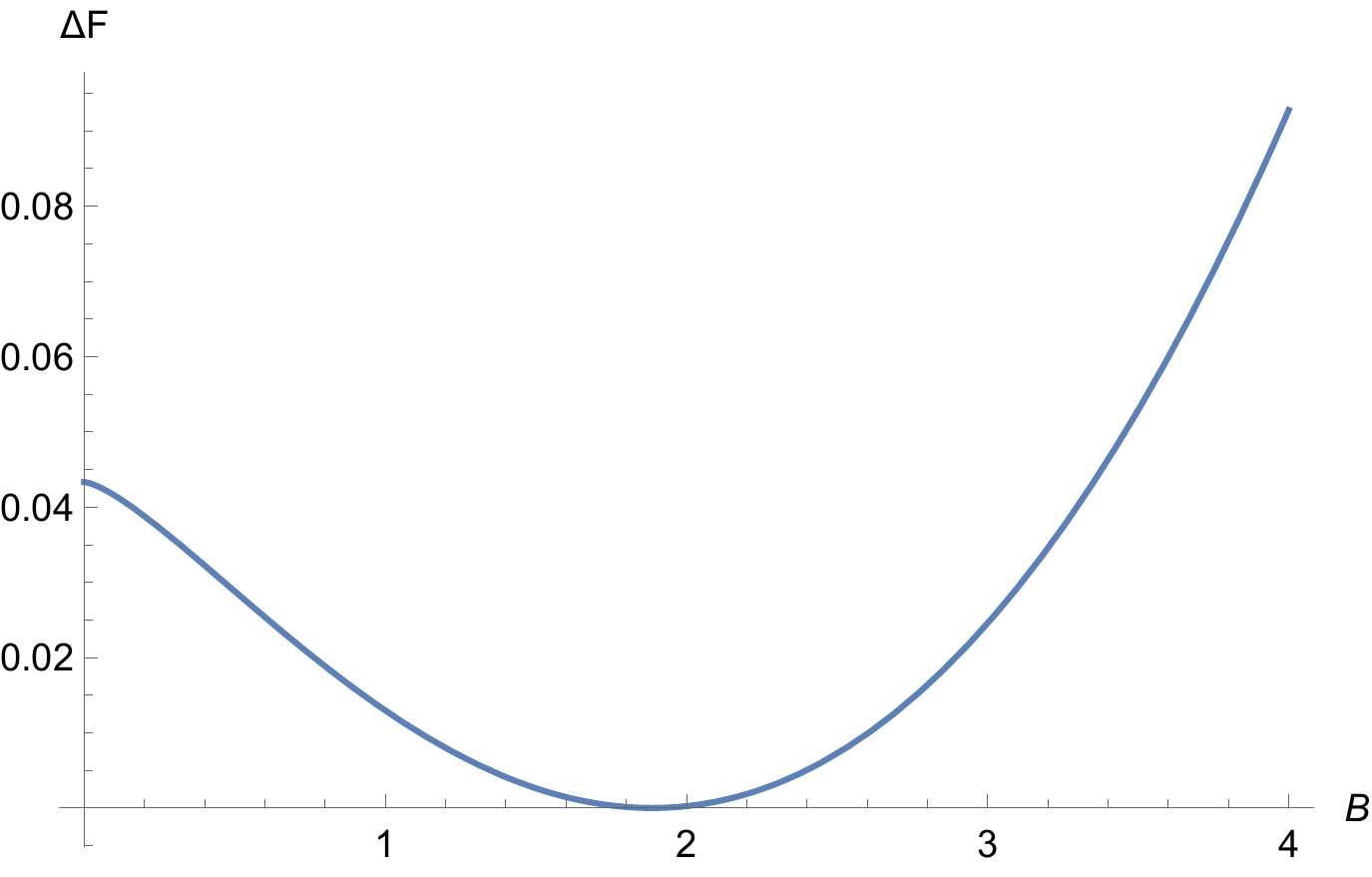}
 \end{center}
 \caption[]{Plot of the difference of the the two free energies for $\chi =1$, as function of $B$.}
 
\label{fig3}
\end{figure}

\subsection{Similarities with the Nambu-Jona-Lasinio model}
\lab{NJL}

Our results may find interesting applications in particle physics, regarding dynamical mass generation and spontaneous flavor symmetry breaking in 2+1 dimensional gauge theories under external magnetic fields (see \cite{Miransky:2015ava} and the references therein). It is well known that magnetic field acts as a catalyst of chiral condensate in 3+1 dimensional gauge theories with massless fermions \cite{Gusynin:1994re,Gusynin:1995nb}. 
As shown in  \cite{Pisarski:1984dj,Appelquist:1986fd}, it also acts as a catalyst for the flavor symmetry breaking $U(2)\to U(1)\times U(1)$ in similar 2+1 dimensional gauge theories, with fermions in a 4-component reducible Dirac representation. These theories are generalized and extensively studied in vector like, 
large-$N_f$ Nambu-Jona-Lasinio (NJL) models, as reviewed in \cite{Miransky:2015ava}. There, the spontaneous symmetry breaking pattern becomes $U(2N_f)\to U(N_f)\times U(N_f)$. 
We observe that our results for the scaling of the condensate in the two phases \lq{}\lq{}BB\rq{}\rq{} and  \lq{}\lq{}TG\rq{}\rq{}, given by equations (\ref{bsol}) and (\ref{bsoltg2}), are in striking similarity with the results obtained in these effective models \cite{Miransky:2015ava}. 

In the absence of magnetic fields, the flavor condensate in NJL models vanish when the quartic fermion coupling $g$ is smaller than a critical value $g_c$. For $g>g_c$ a fermion mass term is dynamically generated and the condensate becomes nonzero \cite{Miransky:2015ava}. Its strength  $\sigma_0$ is proportional to the difference $g-g_c$. This breaks the flavor symmetry as described above. Now we would like to compare this with our values for the condensate in the two phases (\ref{bsol}) and (\ref{bsoltg2}) at $B=0$. First of all, we see that the condensate vanishes in the TG phase, therefore the phase with $g>g_c$ of the NJL model could only be identified with the black brane phase: 
\be\lab{bB0}
\sigma_0 = b_{BB}\bigg|_{B=0} = \sqrt{\frac32}\frac{|\chi|}{2}\, .
\ee
It is therefore tempting to identify the chemical potential $\chi$ with the difference of the 4-fermion coupling and the critical coupling, i.e. $g-g_c$, for $g>g_c$. As explained in \cite{Miransky:2015ava}, two qualitatively different phases arise when $B$ is turned on. In the phase analogous to our black brane phase, the condensate scales as 
\be\lab{njl1}
\sigma_{NJL,1} \approx \sigma_0\le(1 + \frac{B^2}{12 \sigma_0^4}\ri)\, ,
\ee
for $B\ll \sigma_0^2$, which qualitatively agrees with the scaling we have found in (\ref{bsol}): 
\be\lab{bbb}
\sigma_{BB} \equiv b_{BB}\bigg|_{B\neq 0}=  \sigma_0\le(1 - \frac{B^2}{ 384 \sigma_0^4}\ri)\, ,
\ee
which is valid for any value $B$. The second phase is obtained in the region  $g\to g_c$ which corresponds to $\chi/\sqrt{B}<< 1$ limit. In this limit the TG phase definitely wins over the BB phase, as can be seen from  (\ref{fdif}), which is the only phase where scaling of the condensate becomes independent of $\chi$, (\ref{bsoltg}) 
\be\lab{btg}
\sigma_{TG} \equiv b_{TG} = 0.297 \sqrt{B} \, ,
\ee
whereas the NJL model result is 
\be\lab{njl2}
\sigma_{NJL,2} = 0.446 \sqrt{B} \, ,
\ee
again, in qualitative agreement. We note that this qualitative agreement is non-trivial, for it cannot be deduced only by dimensional analysis, as the condensate in our case could have scaled with an arbitrary power of the ratio $B/\chi^2$. 

\section{Fluctuations} 
\lab{flucs}

Another support to our findings, namely presence of quantum criticality at the locus $B~=~B_c(\chi)$, comes from the study of the theory spectrum. 

Let's consider then the spectrum of fluctuations obtained by acting with a bosonic operator ${\cal O}_{\Delta}$ on the vacuum. This can be determined holographically, by studying the fluctuations of the dual bosonic bulk field with mass $m^2 = \Delta(3-\Delta)$, on the gravity background corresponding to the field theory vacuum. 

Below we consider the special case of fluctuations with $m^2=0$,  both on the thermal gas (TG) and the black brane (BB) backgrounds. The spectrum is given by solutions corresponding to energy eigenvalues $\omega$, which are normalizable both in the ultra-violet ($r\to\infty$) and in the infra-red regime (namely $r\to r_s$ on the TG solution, and $r\to r_h$ on the BB solution).  For simplicity we set $\vec{k}=0$ in the following. The fluctuation equation can be obtained from the action
\be\lab{acfluc} 
S_{fluc}  = \int d^4x \sqrt{-g} g^{\m\n} \6_\m \phi\6_\n \phi^* = \int dr d^3x \sqrt{-g} \le\{g^{rr} |\6_r \xi_\o(r)|^2 + \o^2 g^{tt} |\xi_\o(r)|^2 \ri\}\, ,
\ee
where we set $\phi(r,x) = \xi_\o(r) e^{-i\omega t }$. The first term above can be removed by integration by parts and renormalizing away the boundary term \cite{Klebanov:1999tb}, the second term is required to be finite for finite energy fluctuations. Thus, the spectrum is obtained by solving the fluctuation equation 
\be\lab{fluceq} 
\xi_\o''(r) + \frac{d}{dr} \log(\sqrt{-g} g^{rr}) \xi_\o'(r) - \o^2 g^{tt}g_{rr} \xi_\o(r) = 0\, ,
\ee  
and requiring 
\be\lab{acfluc1} 
\lim_{r_{UV}\to\infty} \o^2 \int_{r_{UV}}^{r_{IR}} dr \sqrt{-g}  g^{00} |\xi_\o(r)|^2 < \infty \, ,
\ee
on the solution. Here  $r_{UV}$ denotes a UV cut-off and $r_{IR}$ is either of $r_s$ or $r_h$ depending on the background. Potential divergences in the integral above arise both in the UV and in the IR. For backgrounds with no horizon, requirement of square integrability in these two limits typically results in a discrete spectrum $\o = \o_n$, see for example \cite{Gursoy:2007er}. This conclusion does not necessarily hold for backgrounds with regular horizon and such cases should be studied separately. 

\subsection{Spectrum in the thermal gas phase}

We are going to address the question whether there exists normalizable solutions (according to the normalizability requirement above) with energy $\o$ arbitrarily close to 0. If such solutions can be found, then there is a continuum of states starting just above the vacuum 
$\o=0$. For $\o\ll 1$ the solution can be obtained perturbatively as $\xi_\o(r) = \xi_0 + \o^2 \delta\xi_\o + {\cal O}(\o^4)$. Since the expression in (\ref{acfluc1}) is already quadratic in $\o$ we can safely drop the second term and consider solutions to (\ref{fluceq}) with $\o=0$. The solution can be obtained analytically in this limit as 
\be\lab{zerosol}
\xi_0(r) = C_1 + C_2 \int_\infty^r \frac{dr'}{\sqrt{-g} g^{rr}} = C_1 + C_2 \int_\infty^r \frac{dr'}{r^{'2} f(r')}\, ,
\ee 
where $C_{1,2}$ are integration constants and we used the ansatz (\ref{metric}). The solution with $C_1\neq 0$ is not normalizable near the UV because the limit in (\ref{acfluc1}) diverges linearly in $r_{UV}$. Therefore we set $C_1=0$.   
This solution with $C_2\neq 0$ is clearly normalizable in the UV as the result of the integration in (\ref{acfluc1}) is 
proportional to $r_{UV}^5$. 

Now let us look at what happens near the IR. For $b>0$ the singularity is given by $r_s = 3b$. The TG solution with a good type of singularity at that point is given by setting the parameters as (\ref{good}) in (\ref{metric}). One then shows that
\be\lab{integrand} r^2 f(r) = (r-3b)((r+b)^3 - B^2/(2b)) = 48b^2\eps^2 + {\cal O}(\eps^3)\, ,\ee
where we set $r = 3b + \eps$ and expanded near $\eps = 0$ in the second equation.  To obtain this double zero it is crucial to use the second condition in (\ref{good}), namely $b = 2^{-7/4} \sqrt{|B|}$. Otherwise one obtains a single zero: $f \to {\cal O}(\eps)$.  Therefore the solution behaves as $\xi_0 \sim \eps^{-1}$ near the singularity and the integral in (\ref{acfluc1}) diverges as $\eps^{-2}$. We conclude that {\em there are no normalizable excitations with arbitrarily small $\o$ in the TG phase.}

It is instructive to consider what would happen had we released the condition $b = 2^{-7/4} \sqrt{|B|}$ on the TG solution. If we keep $b$ arbitrary then the blackening factor has a single zero as mentioned above. Then the solution would behave as $\xi_0 \sim \log\eps$ and the integral in (\ref{acfluc1}) would have a finite limit as $\eps \log^2(\eps)$ plus a constant. Thus, one would have obtained a normalizable excitation with an arbitrarily small energy. This should not happen for consistency of the entire picture and we learn that the condition (\ref{good}) is essential for this. 

It is also instructive to carry out the analysis above by releasing our assumption $\o\ll 1$. In this case it is simpler to first consider normalizability in the IR. The solution to the fluctuation equation (\ref{fluceq}) near $r=r_s$ can be obtained analytically in terms of Bessel functions as 
\be\lab{solIR} 
\xi(r) = C_1 (r-r_s)^{-\frac12} J_1\le(\frac{\o_B}{\sqrt{r-r_s}}\ri) + C_2 (r-r_s)^{-\frac12} Y_1\le(\frac{\o_B}{\sqrt{r-r_s}}\ri)\, ,
\ee
where we defined the combination 
\be\lab{defwB}
\o_B = 2^{\frac78}\o/(3B^{\frac12})\, .
\ee
 Requiring normalizability\footnote{Normalizability condition is determined by passing to the Euclidean solution with the Euclidean frequency related to the Lorenztian one as $\omega_E = -i \o$.} in the IR then sets $C_2= i C_1$.  This solution should generically go over to a solution near the UV that is given by the sum of the two independent solutions $\xi \sim c_1$ and $\xi \sim c_2 r^3$ where $c_2$ and $c_1$ are related since we already fixed one of the integration constants by setting $C_1=0$. Therefore this solution is non-normalizable in general, it would be normalizable only at certain discrete values of $\o_n$ that are non-vanishing. We reach the same conclusion that {\em there exist no normalizable excitations with arbitrarily small $\o$ in the TG phase.}

\begin{figure}[t!]
\begin{center}
\includegraphics[scale=0.7]{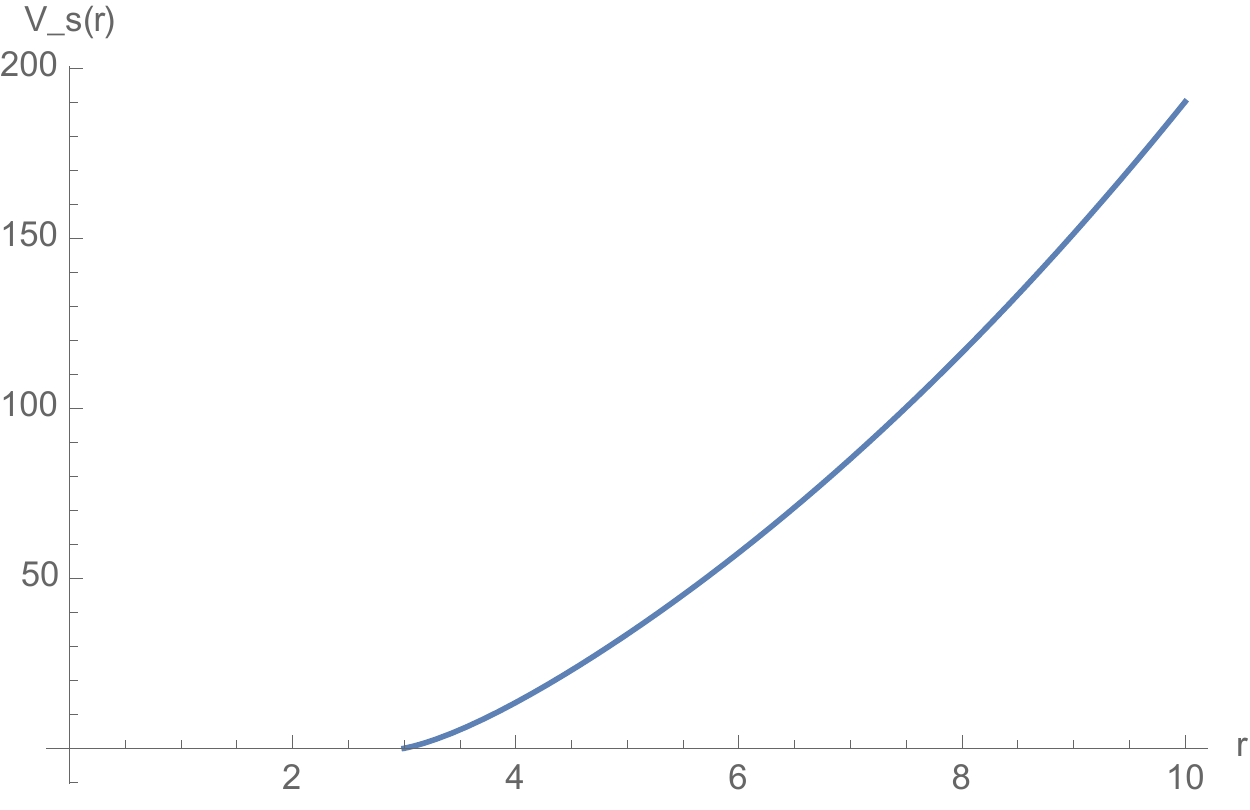}
 \end{center}
 \caption[]{We plot the Schrodinger potential for fluctuations around the thermal gas solution for the choice of $b=1$. The potential vanishes as $r$ approaches the singularity located at $3b$. Therefore the potential has the same qualitative features as the $-g_{tt}$ factor plotted in the right figure in \ref{figpotential}.}
\label{figschro}
\end{figure}

In passing, we note that the fluctuation equation (\ref{fluceq}) can be put in a Schr\"odinger form 
\be\lab{schro} 
-\frac{d\Psi(z)}{dz^2} + V_s(z) \Psi(z) = \omega^2 \Psi(z)\, ,
\ee 
by transforming to the  tortoise coordinate $z$ and making the field redefinition,
\be\lab{defsschro} 
z = -\int_r^{\infty} \frac{dr}{g_{tt}(r)}\, , \qquad \xi_\o = \le(g^{00}g^{rr}\sqrt{-g}\ri)^{-\frac12}\Psi\, ,
\ee
where $g_{\m\n}$ is the metric of the thermal gas solution given by (\ref{mettg}). Luckily one obtains an analytic expression for the Schr\"odinger potential in the $r$-variable as
\be\lab{schropot}
V_s(z) \equiv \tilde{V}_s(r) = \frac{(3 b-r) \left(21 b^2+6 b r+r^2\right) \left(279 b^4-54 b^3 r+b^2
   r^2-4 b r^3-2 r^4\right)}{(b+r)^5}\, .
\ee 
We plot this potential for a choice of $b=1$ in figure \ref{figschro}. Note that the potential vanishes as $r$ approaches the singularity located at $r=3b$. This does not mean however that there exists massless states: as we have shown above states with $\o=0$ does not have square-normalizable wave-functions. We also note that the Schr\"odinger potential enjoys the same qualitative features with the $-g_{tt}$ factor plotted in the left figure in \ref{figpotential}.

\subsection{Spectrum in the black brane phase}

In contrast to the thermal gas phase above, one can show that there exist normalizable fluctuations with arbitrarily small frequency $\o$ in the black brane phase. The easiest argument is as follows. The analog of the normalizable modes in the black brane with Lorentzian signature are the quasi-normal modes (QNM). They are fluctuations on the black brane background with infalling boundary conditions on the horizon and vanishing Dirichlet boundary condition on the boundary.  At finite $T$, the QNM spectrum is typically given by separated poles on the lower complex frequency plane, the lowest QNM having $|\o|\sim T$, therefore one can think of these fluctuations as gapped, see for example the review paper \cite{Berti}. If we view the extremal brane as the $T\to 0$ limit of a finite $T$ black brane background, then we indeed find modes with arbitrarily small energy in the spectrum \cite{Starinets:2002br}\cite{Edalati:2010hk}. This argument is robust as long as one keeps an arbitrarily small but finite $T=\eps$  as an IR cut-off. Then the lowest QNM  indeed has  an arbitrarily small energy $|\o|\propto \eps$ and the separation between the QNMs are also of the same order, $|\Delta \o| \sim \eps$. In the strict $T\to 0$ limit however, multiple QNMs accumulate at the origin of the $\o$ complex plane producing a branch-cut \cite{Starinets:2002br}\cite{Edalati:2010hk}. This is in accordance with the holographic correspondence as one expects branch-cuts in the retarded Green's function in a strongly interacting conformal field theory at vanishing temperature. 

\subsection{Extended probes}
\lab{extended}
 
We have also studied the solutions of string-like extended objects in our backgrounds. We summarize here our findings; the details are presented in Appendix \ref{appB}.  It is important to study extended objects on our backgrounds because the corresponding field theory quantities such as the {\em Polyakov loop}, the {\em Wilson loop} and the {\em entanglement entropy} may potentially be considered as order parameters of the critical behavior we found in section \ref{Thermo}. 

In section \ref{appB1} we calculate the action of a Nambu-Goto string wrapped on the Euclidean time direction and extending in the holographic direction from the boundary to the origin in our geometries. We studied this on both of the backgrounds where the origin corresponds to the horizon in the case of the Euclidean black brane, and to the singularity $r_s$ in the case of the thermal gas.  the exponential of this string action corresponds to the Polyakov loop in the corresponding field theory \cite{Witten:1998zw}. We find that, with a proper renormalization at the boundary, the area swiped by the string is finite on both backgrounds, whereas it should be infinite on the thermal gas and finite on the black brane, had these two geometries corresponded to confined (deconfined) phases of the corresponding field theory. 

Secondly, in section \ref{appB2} we studied the string that is attached on the boundary at two points $-l/2$ and $+l/2$ on the x-axis, and hanging down towards the origin. Action of this string corresponds to the potential $V_{q\bar{q}}$ between a quark-antiquark pair  located at $x=-l/2$ and $+l/2$ \cite{Maldacena:1998im,Rey:1998ik}. Thus, this potential should grow with $l$ in a confining phase. We find however that 
on the thermal gas phase the $V_{q\bar{q}}$ remains finite and it approaches to a constant as $l$ is increased. Therefore this phase does not exhibit a confining behaviour. 

The entanglement entropy of a region $A$ with its complement on the boundary theory is obtained in the holographic dual by studying the area of a space-like minimal surface that ends on the boundary of region $A$ \cite{Ryu:2006bv}. This quantity  may also act as an order parameter in confinement-deconfinement type transitions \cite{Klebanov:2007ws}. We investigate this possibility in Appendix \ref{appB3} where we study the space-like minimal surfaces on the thermal gas background. We find that the connected surfaces always have an area smaller than the corresponding disconnected surfaces with the same boundary conditions. This means that the thermal gas, as the black brane, always correspond to a deconfined state in the corresponding field theory.  

We found that none of these quantities provide an order parameter for the phase transition, as they exhibit the  same qualitative behaviour both on the black brane and the thermal gas phases. On the other hand, this negative statement provides a valuable insight on the nature of the critical point, namely that it {\em does not correspond to a confinement-deconfinement type transition}.

\section{Discussion and Outlook} 
\lab{Discussion}

Our main result is a second order phase critical behaviour in the free energy at vanishing temperature, between an electrically and magnetically charged black brane and a magnetically charged, horizonless thermal gas solution, in the Einstein-Maxwell-scalar theory defined by the action (\ref{action}). Both geometries asymptote to  $AdS_4$ near the boundary, and we work in a mixed ensemble where the magnetic charge $B$ and the electrostatic potential $\chi$ are held fixed, and the temperature $T$ is set  to zero. Quite conveniently, both the backgrounds and the physical quantities such as the thermodynamic potentials can be obtained analytically in our study. Moreover, the action can be embedded in M-theory \cite{Cvetic:1999xp}. The critical point we find is somewhat trivial from the gravity point of view, as we {\em define} the thermal gas solution by a limiting procedure where the horizon is sent to the singularity. It is then obvious that these two solutions become the same in this limit. However, it is quite non-trivial from the boundary field theory point of view, as, when expressed in terms of the physical variables, $\chi$ and $B$ the critical behavior takes place at a finite value of $B$ and $\chi$ and $T=0$. Hence, it corresponds to a line of quantum critical points in the phase space, that can be parametrized by $\chi$ or $B$, as in equation (\ref{pt}).     

Our dual field theory is a 2+1 dimensional, strongly coupled gauge theory related to the  deformed ABJM model \cite{Aharony:2008ug} as explained in section \ref{Intro}. From a bottom-up perspective we have a nonconformal, strongly coupled gauge theory in 2+1D at finite chemical potential $\chi$ and magnetic field $B$, holographically defined by the gravity solution. The gauge group $U(1)^2$ in the gravitational theory  corresponds to part of the global R-symmetry of the boundary field theory, and this group is weakly gauged to produce background magnetic and electric fields. The theory is also deformed by a bosonic triple trace operator\footnote{As discussed in footnote \ref{bc-choice}, we interpret the scalar field as having mixed boundary conditions. Other choices are possible for the bulk solution under consideration, leading to different holographic interpretations that are not of interest in this work.}---that corresponds to the bulk scalar $\varphi$--- whereby breaking the conformality of the theory and initiating an RG flow. The theory is non-conformal even at vanishing $B$ and $\chi$. This is clear from the running of the scalar $\varphi$ when $b\neq 0$. Therefore we are dealing with a non-conformal strongly coupled gauge theory that contains both fermionic and bosonic fields in the adjoint representation of the gauge group. It is important to note however, that the deformation of the field theory by the operator $\Phi^3$ is not relevant, in fact it is classically marginal, since the engineering dimension of this operator is 3 by dimensional analysis. It is also important to note that the source corresponding to the the operator $\Phi$ is set to zero in the field theory, see the discussion on page 12 of \cite{Papadimitriou:2007sj}. Therefore the only dimensionful scales in the theory, at vanishing temperature, are the electrostatic potential $\chi$ and the magnetic field $B$.  

Below we discuss the implications of our findings in the holographic dual field theory and in supergravity.  

\paragraph{Nature of the quantum critical point:} 
 First of all, as we showed in section \ref{extended} by studying the Polyakov loop, the quark potential and the entanglement entropy that the quantum critical point we find here does {\em not} correspond to  a confinement-deconfiement type transition. However, it may correspond to spontaneous breaking of a flavor symmetry in the dual field theory.  We already noted similarities between our findings and the earlier studies in 2+1 dimensional Nambu-Jona-Lasinio models in section \ref{NJL}. In particular we observed that the dependence of the VeV of the scalar operator follow a similar pattern observed in these studies: there is  agreement between the square-root scaling in the  TG phase and the near critical region of the NJL model, as well as between the scaling in the BB phase and the $g>g_c$ region of the NJL model. There is one important difference in the latter case however. Comparison of equations (\ref{njl1}) and (\ref{bbb}) shows that while the magnetic field tends to increase the value of the condensate in the NJL model, it tends to destroy the condensate in our holographic model.  The former phenomenon is called the {\em magnetic catalysis} \cite{Gusynin:1994re,Gusynin:1995nb}, a phenomenon well-established by perturbative and effective field theory calculations (see \cite{Miransky:2015ava} for a recent review, see also a study of the phenomenon in the ABJM type models in \cite{Jokela:2013qya}). Decrease of the condensate with $B$ in confining 3+1D gauge theories above the deconfinement temperature has also been observed both on the lattice \cite{Bali:2012zg}, in the Sakai-Sugimoto model \cite{Preis:2010cq,Preis:2012fh} and in the hard-wall model \cite{Mamo:2015dea}, and it was termed the {\em inverse magnetic catalysis}. The same phenomenon also occurs in 2+1D, as studies on the lattice, the NJL model and the NJL with Polyakov loop shows \cite{Ferreira:2014kpa}. What we find here  is the  inverse magnetic catalysis at vanishing temperature in a holographic dual of a strongly interacting gauge theory. 
One should understand this phenomenon from a microscopic  point of view. We suspect that both strong interactions and non-conformality of our theory  is  essential in this respect. Firstly, as one can show in perturbative studies quite generally, there is no inverse magnetic catalysis when the interactions are weak \cite{Gusynin:1994re,Gusynin:1995nb}. Secondly, even the strongest interactions cannot allow for a condensate, hence a fermion mass term, when the underlying theory is conformally invariant. 

\paragraph{Implications for the ABJM-type models:} Regardless of the discussion above, it would be interesting to investigate possible patterns of flavor symmetry breaking in the ABJM model under magnetic field directly from the microscopic point of view. As discussed in \cite{Aharony:2008ug} the superpotential enjoys a full $SU(4)$ flavor symmetry obtained by combining the $SU(2)\times SU(2)$ symmetry that rotates the $A$ and $B$ type superfield doublets separately, and the $SU(2)_R$ R-symmetry. The breaking $SU(4)\to SU(2)\times SU(2)$---that is very similar to the pattern in the NJL model above---would take place through  formation of a condensate in a non-conformal cousin of ABJM. This can happen either spontaneously through strong interactions, or explicitly with aid of an external magnetic field. All of these questions should first be studied in the perturbative ABJM model.  

When the $\mathcal N=2$ Supergravity theory considered in this paper is embedded in $\mc N=8$ Supergravity (as required if one is interested in ABJM as the dual theory), the stability of the gravity configuration needs to be discussed taking into account the full $\mc N=8$ theory. In particular, it was shown in \cite{Donos:2011pn} that the $AdS_2\times \mathbb R^2$ vacua are unstable in the $\mc N=8$ theory for generic values of the parameters. These instabilities arise because charged scalars of the $\mc N=8$ theory do not satisfy the $AdS_2$ BF bound. Notice that, in our analysis, also the supersymmetric configuration is thermodynamically disfavoured with respect to the thermal gas, at $T=0$. This can be interpreted in the dual theory as the supersymmetric vacuum being disfavoured with respect to a scalar field condensate. In order to understand the BPS magnetic solution in the context of ABJM one should analyze the instability of such specific background along the lines of \cite{Donos:2011qt,Donos:2011pn,Almuhairi:2011ws}. This will be left for future work.

\paragraph{Applications to condensed matter theory:} Despite its exotic nature, the ABJM theory at finite charge density comes very close to various realistic applications in  2 (spatial) dimensional condensed matter  systems such as the cuprate superconductors and the various strongly coupled semimetals, see \cite{Hartnoll:2009sz,McGreevy:2009xe,Herzog:2009xv, Sachdev:2010ch} for reviews. The field theory we consider in this paper is a non-conformal, strongly coupled field theory related to ABJM in the manner explained in section \ref{Intro}. When put under external  magnetic fields, this theory is quite interesting in view of condensed matter applications. It is long argued that resolution to the various puzzles concerning high $T_c$ superconductivity may be associated with presence of a quantum critical point under the superconducting dome \cite{Sachdev}. Connections  between superconductivity \cite{Miransky:2015ava} and spontaneous symmetry breaking in the NJL type effective theories are also well-known, and our observations above may be interesting from this point of view.  All in all, it remains to be seen whether the quantum criticality we found may serve as a proxy for a strongly interacting quantum critical point that may underlie high $T_c$ superconductivity.  To explore this issue it is crucial to study our system at finite $T$. It is also crucial to study correlators of e.g. composite fermionic operators and explore their behaviour near the quantum critical point. We plan to return these questions in the near future.

Quite interestingly, there also seems to be an inherent connection between the spontaneous breaking in the NJL models above and the studies of graphene at strong Coulomb interactions   \cite{Khveshchenko:2001zza}. In the theory of graphene, the flavor symmetry we describe in section \ref{NJL}  may be viewed as a chiral symmetry of a 4-component Dirac fermion constructed out of fermionic excitations around two conical points and two sublattices in the bipartite hexagonal lattice of a 2D graphene sheet. Spontaneous symmetry breaking and dynamical generation of a fermion mass has the effect of a semimetal-insulator transition at vanishing temperature \cite{Khveshchenko:2001zza}, which has been observed in  highly oriented pyrolitic graphite in the presence of magnetic field perpendicular to the layers \cite{Kopelevich}. Can we view the critical behavior we find here as a similar  semimetal-insulator type transition? It would be interesting to answer these questions by calculating thermal and electric conductivities in our holographic model, and this is in fact an issue we plan to investigate in the near future. 

Finally, it is very interesting to explore the fate of the phase that we found in this paper in the regime of finite temperature. In particular we would like to know if there exists a phase separation line in the $(T,B)$ plane, at fixed $\chi$, of second or higher order phase transitions that ends on the quantum critical point in the vanishing $T$ limit. It is also conceivable that the critical behavior we find extends into a true second order phase transition at finite $T$ domain. We plan to address these questions in a future work. Also, the precise shape of the crossover lines that separate the critical and non-critical regions shown in figure  \ref{fig1} is a question to be addressed in future work. 

\paragraph{Open questions in the Supergravity context:} Last but not least, our study raises interesting questions directly in the context of supergravity.  
First of all, our ensemble includes BPS solutions \cite{Cacciatori:2009iz}. One may wonder how the BPS brane decays into the thermal gas solution, despite being stable with respect to small fluctuation in charge\footnote{This can be done, for instance by computing the permittivity of the solution, defined as $$ \varepsilon= \left( \frac{\partial Q}{\partial \phi }\right)_T\,.$$ For the BPS configuration this quantity is positive, denoting stability with respect to charge fluctuations: the chemical potential of the black brane increases as a result of placing more charge on it.}. In general, supersymmetric solutions saturates a BPS bound of the form $M\geq |Q|$. For zero temperature solutions, the mass coincides with the system's free energy in the canonical ensemble. Thus, one explains the stability of the supersymemtric solution in the thermodynamic phase space. We considered however a \textit{mixed canonical-grand canonical} ensemble, where the free energy is given by \eqref{fbb1}, thus saturation of the BPS bound is no more equivalent to non-perturbative thermodynamic stability. With the choice of this ensemble, we indeed find that in most of the parameter space except the locus of critical point, the thermal gas is thermodynamically dominant at vanishing temperature. 

It is interesting to ask if the thermal gas itself preserves any supersymmetry, and if so if it preserves more supersymmetries than the black brane. In this case, the thermodynamically favoured solution would be the most supersymmetric one. 
In order to address this question, one has to embed the theory we considered in $\mathcal{N}=2$ $U(1)$-gauged Supergravity (a truncation of $\mathcal{N}=8$ $SO(8)$ gauged theory where only the diagonal $U(1)$ in the Cartan of $SO(8)$ is gauged). Here, however, an electric-magnetic duality transformation affects not only the Maxwell fields, but also the gauging parameters. Such a transformation has been performed in \cite{Dall'Agata:2010gj} in the black hole case; the analogue, black brane 1/4-BPS constraint obtained with this transformation is $3B-q=0$. One can then see that the thermal gas, having $q=0$, cannot preserve supersymmetry for $B\neq0$ (we find that the thermal gas cannot be supersymmetric for either choice of Killing spinors studied in the literature \cite{Duff:1999gh,Dall'Agata:2010gj,Hristov:2010ri}).  We cannot exclude less conventional duality transformations on the gauging that acts differently than in \cite{Dall'Agata:2010gj}, but, since the origin of magnetic gauging parameters in M-theory is less understood \cite{Dall'Agata:2012bb}, the analysis of these scenarios goes beyond the scope of the present analysis.

Finally, a pressing question relates to possible resolution of the thermal gas singularities in  the full string theory. It will be very interesting to see if the singularities at $r=3b$ and $r=-b$ can be excised through an enhan\c{c}on mechanism found in the study of D1-D5 and D2-D6 systems, see \cite{Johnson:1999qt}. 

\section*{Acknowledgements}

It is a pleasure to acknowledge interesting discussions with Nikolay Bobev, Adam Bzowski, Nele Callebaut, Frederik Denef, Federico Galli, Steven Gubser, Gavin Hartnett,  Ioannis Papadimitriou, Christopher Rosen and   Stefan Vandoren.  

A.G. acknowledges support by the FWO - Vlaanderen, Project No.
G.0651.11, by the Interuniversity Attraction Poles Programme
initiated by the Belgian Science Policy (P7/37) and by the FWO Odysseus program. C.T. acknowledges support from the NWO Rubicon Grant and DOE grant DE-SC0011941.
This work has been supported in part by COST Action MP1210 The String Theory Universe. This work was
supported by the Netherlands Organisation for Scientific
Research (NWO) under VIDI grant 680-47-518, and the
Delta-Institute for Theoretical Physics (D-ITP) that is
funded by the Dutch Ministry of Education, Culture and
Science (OCW).

\begin{appendix}

\section{On-shell action via holographic renormalization and background subtraction}

\subsection{On-shell action via holographic renormalization}\lab{appA1}

In this section we compute the on-shell action 
for our configurations.
In doing so, we plug in the action \eqref{action} the solutions of the equations of motion and perform the integration over all space. Since the quantity $S_{on-shell}$ obtained this way is per se divergent, we need to resort to the techniques of holographic renormalization (see for instance \cite{deBoer:1999tgo,deHaro:2000vlm,Bianchi:2001kw,Papadimitriou:2005ii}). We therefore add to the action \eqref{action} appropriate counterterms that we denote by
\be \label{stotonshell}
S_{on-shell, ren} = S_{on-shell} + S_{ct} \,.
\ee
The counterterms we need to add are functions of boundary curvature invariants, hence they do not alter the bulk equations of motion. The prescription for such terms is spelled out for example in \cite{Papadimitriou:2007sj} (see also \cite{Batrachenko:2004fd} for a related computation). They are constructed from the boundary Ricci scalar $\mathcal{R}_3$ and a function $\mathcal{W}(\varphi)$ of the scalar fields, called superpotential:
\begin{equation} \label{count_can}
S_{ct}=  \frac{1}{4 \pi G} \int_{\partial M} d^3x \sqrt{h} \left( \, \frac{\ell_{AdS}}{2} \mathcal{R}_3-\mathcal{W}(\varphi)   \right)\,.
\end{equation}
In this expression, $h$ is the determinant of the induced metric at the boundary $\partial M$, $h_{ab}$, and $\mathcal{R}_{3,ab}$ is its Ricci curvature. For the black brane the curvature of the boundary is zero, hence the first addendum vanishes identically. The superpotential $\mathcal{W}$ appearing in \eqref{count_can} satisfies the following relation:
\be\label{superp}
V_g(\varphi) = \frac12 \left(-\frac32 \mathcal{W}^2 + g^{ij} \partial_i \mathcal{W} \partial_j \mathcal{W} \right)\,.
\ee
The crucial point in the computation of counterterms for mixed boundary condition is the fact that the superpotential relation \eqref{superp} fixes the function $\mathcal W)(\varphi)$ up to finite terms, that would in principle affect the renormalized physical quantities. Indeed, to completely fix this finite term one needs an additional requirement, namely that the holographic renormalization be derived via a well-defined variational principle. As explained in 
\cite{Papadimitriou:2007sj}, only one specific choice of finite term (thus of $\mc W$, satisfying \eqref{superp}) satisfies this requirement\footnote{It has been noticed in \cite{Gnecchi:2014cqa} that, when the boundary conditions enforce a marginal deformation, the finite part of the counterterm $\mc W$ can be equivalently determined by taking the superpotential $\mc W\equiv \mc W_{flow}$ that drives the flow of the scalar fields, $ \varphi' = \partial_{\varphi} \mathcal{W}_{flow}$ of the solution. See \cite{Gnecchi:2014cqa} for more details.}. Following the prescription of \cite{Papadimitriou:2007sj} and the work \cite{Gnecchi:2014cqa} we find that the correct superpotential counterterm one has to use for our solutions (both branes and thermal gas) is
\begin{eqnarray}\label{Wct}
\mathcal W(\varphi)&=&\frac2{\ell_{AdS}}\left( 1+\frac{\varphi^2}{2}+\frac{1}{6\sqrt6}\varphi^3+\mc O(\varphi^4)\right)\ ,
\end{eqnarray}
the finite part being given by the $\varphi^3$ term.

The action obtained here requires specific boundary conditions on the vector fields: these are exactly imposing fixed electric chemical potential and fixed magnetic charge. Let us mention the fact that imposing fixed electric charge would amount in adding the finite Hawking-Ross boundary term \cite{Hawking:1995ap}
 \be
S_{HR}= \frac{1}{4 \pi G} \int_{\partial M} d^3x \sqrt{h} \, n_{a} F^{ab} A_{b}\,,
\ee
where$n_a$ is an outward pointing vector normal to the boundary. 
Since we have decided to work in the mixed ensemble (with fixed electric chemical potential and fixed magnetic charge) we do not need to make such an addition.

We are now ready to perform the computation of the on-shell action. We first start by noticing that, making use of the Einstein's equations of motion, the 
action \eqref{action} can be rewritten in terms of the Ricci tensor components plus the integrals over the kinetic terms of the gauge fields:
\begin{equation}\label{rewriting}
S_{on-shell}=\frac{\mathcal{V} \beta}{4 \pi}\int_{r_{h}}^{\infty} dr \sqrt{-g}\left[(R_{y}^{y}+R_{t}^{t})/2+\mc I_{\Lambda \Sigma}F_{\mu \nu}^{\Lambda}F^{\Sigma\,\mu \nu}\right] + S_{GH} \,.
\end{equation}
We first impose a radial cutoff $r_0$ that should be sent to infinity after integration.  The extrema of integration are $r_h$ and $r_0$.
The total on shell action takes the form
\be \label{onshellact}
S_{on-shell}= \frac{\mathcal{V} \beta}{8 \pi} \left(-\frac{3}{4} \left(2 c_1-8 b^3\right)-\frac{3 c_1 b+c_2}{4 (r_h-3 b)}-\frac{3 c_1 b -3 c_2}{4 (r_h+b)} +3 b^2 r_h-r_h^3+6 b^2 r_0-2 r_0^3 \right)\,,\ee
where $\beta$ comes from the integration in the time direction.
For the solution at hand, the computation of the counterterm using \eqref{Wct} gives
\be \label{countterm}
S_{ct}= \frac{\beta \mathcal{V}}{4\pi} \left( \frac{1}{4} \left(2 c_1-8 b^3\right)-3 b^2 r_0 +  r_0^3  \right) \,.
\ee
Plugging in  \eqref{onshellact} and \eqref{countterm} in \eqref{stotonshell} we see that the divergencies cancel, giving a finite result for the on-shell action. Using the horizon equation $f(r_0) =0$, we can recast \eqref{stotonshell} in the following form:
\be \label{ren_osaction}
S_{os,ren}  = \frac{1}{16 \pi}  \beta \mathcal{V} \left(3 (3b - r_h) (b+r_h)^2-2 c_1\right) \,.
\ee
At this point we can directly rewrite the on-shell action in terms of the thermodynamic potentials. The mass of the system can be computed from the renormalized stress energy tensor $\tau^{ab}$
\be
\tau^{ab} = \frac{2}{\sqrt{h}} \frac{\delta S_{ren} }{\delta h_{ab}}\,.
\ee
The following expression gives the conserved charge associated with the boundary Killing vector $K_b$ ($\sigma^{ab}$ is the induced metric on the spacelike section $\Sigma$ of the boundary and $u_a$ is the unit normal vector to $\Sigma$):
\be
Q_{K}=\int_{\Sigma} d^2x \sqrt{\sigma} u_a \tau^{ab} K_b \,. \ee
The mass of the system is then obtained for $K=\partial_t$ and one finds
\be
M = Q_{\partial_t}= - \frac{1}{4 \pi} \frac{c_1}{2}\,.
\ee
This expressions coincides with the mass computed via the AMD procedure \cite{Ashtekar:1984zz,Ashtekar:1999jx}. The temperature $T$ is
\begin{equation}
T = \frac{1}{4 \pi} \frac{df(r)}{dr}\bigg{|}_{r_{h}}\frac{1}{\sqrt{H_{0}(r) H_{1}^{3}(r)}}\bigg{|}_{r_{h}}=\frac{-12 b^2 r_h-c_{1}+4 r_{h}^3}{4 \pi \sqrt{\left(r_h-3b\right) \left(b+r_h\right)^3}}\,,
\end{equation}
and the entropy density reads
\begin{equation}
S = \frac{Area}{4{\kappa}^2} = \frac{r_h^2\sqrt{H_0(r_h)H_1^3(r_h)}}{4{\kappa}^2}\,.
\end{equation}
The magnetostatic potential $m_B$ and the electrostatic one $\chi$ respectively read
\begin{equation}
m_B =  - \int_{r_h}^{\infty} G_{1, tr} dr\,,
\qquad
\chi =  - \int_{r_h}^{\infty} F_{0, tr} dr \,.
\end{equation}
The field $G_{\Lambda}$ is the dual of the field strength $F=dA$ and it is defined in this way:
\begin{equation}
G_{tr, \Lambda} = \frac14 \epsilon_{tr xy } \frac{\partial \mathcal{L}}{\partial F_{xy}^{\Lambda}}= \epsilon_{tr xy} I_{\Lambda \Sigma} F^{xy, \Sigma}
\end{equation}
with Levi Civita tensor 
\begin{equation}
\epsilon_{\mu \nu \rho \sigma} = e_{\mu}^{a}  e_{\nu}^{b}  e_{\rho}^{c}  e_{\sigma}^{d} \epsilon_{abcd}\,, \qquad \epsilon_{0123}=1\,.
\end{equation}
With $\epsilon_{tr x y} = h^2(r) $ and $F_{xy}^{\Lambda}= \frac{p^{\Lambda}}{2} $ one gets 
$
G_{tr, \Lambda}= \frac{I_{\Lambda \Sigma} p^{\Sigma} }{2 h^2}$,
hence
\begin{equation}\label{chi0}
\chi = \frac{q}{2 \left(3 b-r_h\right)}\,, \qquad m_{B} =  \frac{3 B}{2 \left( b+ r_h\right)}\,.
\end{equation}
Having all the conserved quantities and the potential defined, one can check that the first law of thermodynamics is satisfied:
  \begin{equation}
dM = T dS - \chi d q  + {m} d B\,,
 \end{equation}
 and the (renormalized) on-shell action $S_{on-shell,ren}$ coincides with the free energy for the mixed ensemble
\be
\frac{S_{on-shell,ren}}{\beta} = \frac{I}{\beta} = M- TS + \chi q\,.
\ee

\subsection{Background subtraction method}

We will now illustrate the background subtraction method for the free energy computation. The appropriate background should have the same boundary asymptotics of the solution taken into consideration and, in addition to it, it has zero AMD mass.

Obviously, we should subtract the same background from the two solutions that we want to compare. Using different backgrounds would result in a finite piece that comes from the difference between the two backgrounds.

For our specific example, such a background  turns out to be the domain wall solution with metric 
 \begin{equation}\label{our_solution}
 {\rm d} s^2 = -r^2\sqrt{H_0(r)H_1^3(r)} \,{\rm d}t^2 + \frac{ {\rm d}r^2}{r^2\sqrt{H_0(r)H_1^3(r)})} + r^2\sqrt{H_0(r)H_1^3(r)}\, {\rm d} \sigma^2\ ,
\end{equation}
and area element \begin{equation}
{\rm d}\sigma^2= {\rm d}x^2 + {\rm d}y^2\,.
\end{equation}
Furthermore, the background has zero magnetic and electric charges $F_{\mu\nu}=(0,0,0,0)$ and the same asymptotic expansion for the scalar field at infinity. %

To compute the finite on-shell action we subtracts the on-shell actions of the background from the on-shell action of the black brane.  We can use the same rewriting  \eqref{rewriting} used in the previous section. Since for the background configuration the electric and the magnetic charges are zero, the on-shell action can be written as a whole as an integral of the Ricci tensor component $R_{t}^{t}$.
One can then expand the difference around $r_0 \rightarrow \infty$. The result is given in terms of a finite part and subleading terms that go to zero as $r_0\rightarrow\infty $. 
\begin{equation}
I=  \frac{\beta \mathcal{V}}{4\pi} \left( \frac{9 {b}^3+ 15 {b}^2 r_{h}+3 b r_{h}^2-2 c_{1}-3 r_{h}^3}{4} \right)
 +\mathcal{O}({r_0}^{-1})
\end{equation}

This quantity coincides with the renormalized on shell action \eqref{ren_osaction}, and the free energy $F$ is the finite part divided by $\beta$, which is exactly $F=M-TS+q \chi $. Hence in this case the background subtraction gives the same result as the holographic renormalization.

\subsection{On-shell action for the thermal gas solution}
As we show in a previous section the way we acquire the acceptable thermal gas solution is to take the limit $r_{h}\rightarrow r_{s}$. Likewise to compute the on-shell action we will perform the same procedure as in the black brane case but the limits of integration will be $r_{s}+\epsilon ,\, \epsilon\rightarrow 0$ and $r\rightarrow\infty$. After the subtraction of the background one finds
\begin{equation}
F_{sol}=M_{ADM, sol}=- \frac{c_{1}}{8 \pi }\,.
\end{equation}
All the rest thermodynamic quantities are computed as in the black brane but taking the limit $r_{h}\rightarrow r_{s}$.

As one can notice, we have found the thermodynamic quantities and the free energy as functions of $r_{h},c_{1} ,b$, using the $T=0$ condition we can express them in terms of the actual thermodynamical variables $T=0, \chi, B$.

\section{Extended objects}
\lab{appB} 

\subsection{Polyakov loop} 
\lab{appB1}

Action of a Euclidean string propagating on the background is given by 
\be\lab{NG} 
I_{NG} = \frac{1}{4\pi\alpha'} \sqrt{ \textrm{det}(\6_\a X^\m \6_\b X^\n g_{\m\n}) }d\sigma d\tau\, ,  
\ee
where $g_{\m\n}$ is the Euclidean metric of the target space. This area diverges in an asymptotically AdS space-time and the renormalization procedure is standard \cite{Maldacena:1998im}. After renormalization, the area on the Euclidean black brane background is obviously finite because the near horizon geometry is flat corresponding to the origin  in the Euclidean signature. The area on the thermal gas background can be obtained by choosing a gauge $\sigma = r$, $\tau= t_E$ where $t_E$ is the imaginary time. Then one obtains $\textrm{det}(\6_\a X^\m \6_\b X^\n g_{\m\n}) = g_{tt} g_{rr} = 1$ where we used (\ref{metric}) with $t= i t_E$. Thus, after renormalization of the UV divergence, we again find a finite area for the Nambu-Goto string on the thermal gas, even though there is a singularity at $r=r_s$. 

\subsection{The quark potential} 
\lab{appB2}

We consider a string embedded in the thermal gas background that is attached on the boundary at two points separated by a distance $l$.  A typical geometry minimizing the (\ref{NG}) with this boundary condition is a curve that ends on the boundary  $r=\infty$ at the points $(x,y) = (\pm l/2,0)$ and hangs down the interior of the geometry making a turning point at $r=r_0$, see for example \cite{Kinar:1998vq}. The on-shell action is proportional to the quark-antiquark potential on the boundary and the latter can be put in a form\cite{Kinar:1998vq} 
\be\lab{vqq}
V_{q\bar{q}} = H(r_0)\, l - 2d(r_0)\, 
\ee 
where 
\be\lab{defd} 
d(r_0) = 2 \int_{r_0}^{\infty} dr\frac{G(s)}{H(s)} \le(\sqrt{H^2(r)-H^2(r_0)} - H(s) \ri)  - 2\int_{3b}^{r_0} dr G(r)\, ,
\ee
with $ H^2 = g_{tt} g_{xx} $ and $G^2 = g_{tt} g_{rr}$. On our backgrounds these quantities simplify as $G^2(r) = 1$ and $H^2(r) = f(r) r^2$ where $f$ is defined in (\ref{warp-factors}). On a confining background, as $l$ is increased, the turning point of the string, $r_0$ approaches to a final value $r=r_f$ located deeper in the interior of the geometry where $d(r_f)$ and $c(r_f)$ attain finite values. Therefore, one obtains linear confinement in (\ref{vqq}). On our thermal gas geometry this point corresponds to the singularity $r_f=3b$. In order to explore behaviour of the function $H$ near this point we set $r_0=3b + \eps$ and expand for small $\eps$.  The blackening factor $f$ on the thermal gas can be put in the form 
\be\lab{warpfactor-ftg} 
f(r) = H_0(r) \le(r^2 H_1^3(r) - \frac{B^2}{2br}\ri)\, ,
\ee 
where $H_0$ and $H_1$ are defined in (\ref{warp-factors}). Using this expression and the good singularity condition  (\ref{good}) we obtain 
\be\lab{limH} 
H(r_0) = r_0^2 f(r_0) \to 4\sqrt{3} b \eps\, ,
\ee
in the limit $\eps \to 0$.  Therefore, we find that $d(r_0)$ in (\ref{vqq}) remains finite as $r_0\to 3b$ as $l$ is increased, whereas  $H(r_0)$ vanishes linearly in this limit. This means that the only way confinement may arise from (\ref{vqq}) is by $l$ diverging faster than $1/\eps$. The latter is given by \cite{Kinar:1998vq} 
\be\lab{lsol} 
l = 2 \int_{r_0}^{\infty} dr \frac{G(r)}{H(r)}\frac{H(r_0)}{\sqrt{H^2(r)-H^2(r_0)}} = 2 \int_{r_0}^{r_1}\frac{\sqrt{f(r_0)r_0^2}}{f(r)r^2} \frac{dr}{\sqrt{f(r)r^2 - f(r_0) r_0^2}}\, .
\ee
 Changing integration variable $r = 3b + \eps$, $r_0 = 3b + \eps_0$, to focus near the singularity,  
we find
\be\lab{llim} 
l =  \int_{\eps_0}^{\infty} \frac{d\eps}{4\sqrt{6} \sqrt{\eps-\eps_0} + \cdots} \, ,
\ee
that always remain finite. We conclude that the string is not confining on the thermal gas solution. 
\subsection{Entanglement Entropy}
\lab{appB3}

Entanglement entropy can be employed to check if the thermal gas is a confining background. For a confining background there are two possible minimal surfaces with the same end points on the boundary. When confinement occurs the favoured minimal surface is the two disconnected straight lines extending from the endpoints of the line segment inside the bulk. In the deconfined phase the minimal surface is the curved line that connects the two endpoints. In the following we identify: $h(r)^2=r^2\sqrt{H_0(r)H_1^3(r)}$.
We divide the boundary region into two parts $A$ and $B$ ($A$'s complement), where $A$ is defined as: $l/2<x<l/2$ and $0<y<\infty$. By parametrizing this 2-dim surface with $(x,y)$ we can write down the action that should be minimized as 
\begin{equation}\lab{area}
A=L\int _{-l/2}^{l/2}dx h\left(r\left(x\right)\right)^2\sqrt{1+\frac{\left({\partial}_{x}r\left(x\right)\right)^2}{U\left(r\left(x\right)\right)^2 h\left(r\left(x\right)\right)^2}}\ .
\end{equation}
One notices that the Lagrangian does not depend explicitly on $x$, hence the quantity
\begin{equation}\lab{ham}
H=-\frac{h(r)^2}{\sqrt{1+\frac{\left({\partial}_{x}r\right)^2}{U\left(r\right)^2 h\left(r\right)^2}}}\ ,
\end{equation}
is conserved.  If $r^{*}$ is the minimal value of r twith respect to $x$, ${\partial}_{x}r|_{r^{*}}=0$, then
\begin{equation}
H=-h\left(r^{*}\right)^2\ ,
\end{equation}
but $H$ is a constant and always equals $-h\left(r^{*}\right)^2$. Thus we can solve \ref{ham} for ${\partial}_{x}r$,
\begin{equation}
{\partial}_{x}r=U(r)h(r)\sqrt{\frac{h(r)^4}{h(r^{*})^4}-1}\ .
\end{equation}
We can now compute the length $l$ that minimizes the area \ref{area} as a function of $r^{*}$,
\begin{equation}\label{l}
\frac{l}{2}=\int_{r^{*}}^{r^{\infty}}dr \frac{1}{U(r)h(r)\sqrt{\frac{h(r)^4}{h(r^{*})^4}-1}}\ ,
\end{equation}
where $r^{\infty}$ is the UV cut-off. Eliminating $l$ from the \ref{area}, we find for the connected surface
\begin{equation}\lab{con}
A^{con}=\frac{L}{2G_{N4}}\int_{r^{*}}^{r^{\infty}}dr \frac{h(r)^3}{h(r^{*})^2U(r)}\frac{1}{\sqrt{\frac{h(r)^4}{h(r^{*})^4}-1}} \ .
\end{equation}
The area for the disconnected case is simply,
\begin{equation}\lab{dis}
A^{dis}=\frac{L}{2G_{N4}}\int_{r_{0}}^{r^{\infty}} dr \frac{h(r)}{U(r)}\ ,
\end{equation}
where $r_{0}$ is infinitesimally close to the singularity.

Plotting the difference between \ref{con} and \ref{dis} $A^{con}-A^{dis}$ as a function of $l$ for all  values of $B$ and $\chi$ we find that the connected minimal surface is always favoured, leading to the conclusion that thermal gas always corresponds to a deconfined phase. 
\end{appendix}


\providecommand{\href}[2]{#2}\begingroup\raggedright\endgroup


\begin{thebibliography}{10}

\bibitem{Gubser:2000nd}
S.~S. Gubser, \emph{{Curvature singularities: The Good, the bad, and the
  naked}}, {\emph{Adv. Theor. Math. Phys.} {\bf 4} (2000) 679--745},
  [\href{http://arxiv.org/abs/hep-th/0002160}{{\tt hep-th/0002160}}].

\bibitem{Aharony:2008ug}
O.~Aharony, O.~Bergman, D.~L. Jafferis and J.~Maldacena, \emph{{N=6
  superconformal Chern-Simons-matter theories, M2-branes and their gravity
  duals}}, \href{http://dx.doi.org/10.1088/1126-6708/2008/10/091}{\emph{JHEP}
  {\bf 10} (2008) 091}, [\href{http://arxiv.org/abs/0806.1218}{{\tt
  0806.1218}}].

\bibitem{Sachdev}
S.~Sachdev, \emph{{Quantum phase transitions, Cambridge University Press
, 2011}} 
\bibitem{Sachdev:2011wg}
S.~Sachdev, \emph{{What can gauge-gravity duality teach us about condensed
  matter physics?}},
  \href{http://dx.doi.org/10.1146/annurev-conmatphys-020911-125141}{\emph{Ann.
  Rev. Condensed Matter Phys.} {\bf 3} (2012) 9--33},
  [\href{http://arxiv.org/abs/1108.1197}{{\tt 1108.1197}}].

\bibitem{Maldacena:1997re}
J.~M. Maldacena, \emph{{The Large N limit of superconformal field theories and
  supergravity}}, \href{http://dx.doi.org/10.1023/A:1026654312961}{\emph{Int.
  J. Theor. Phys.} {\bf 38} (1999) 1113--1133},
  [\href{http://arxiv.org/abs/hep-th/9711200}{{\tt hep-th/9711200}}].

\bibitem{Gubser:1998bc}
S.~S. Gubser, I.~R. Klebanov and A.~M. Polyakov, \emph{{Gauge theory
  correlators from noncritical string theory}},
  \href{http://dx.doi.org/10.1016/S0370-2693(98)00377-3}{\emph{Phys. Lett.}
  {\bf B428} (1998) 105--114}, [\href{http://arxiv.org/abs/hep-th/9802109}{{\tt
  hep-th/9802109}}].

\bibitem{Witten:1998qj}
E.~Witten, \emph{{Anti-de Sitter space and holography}}, {\emph{Adv. Theor.
  Math. Phys.} {\bf 2} (1998) 253--291},
  [\href{http://arxiv.org/abs/hep-th/9802150}{{\tt hep-th/9802150}}].

\bibitem{Duff:1999gh}
M.~J. Duff and J.~T. Liu, \emph{{Anti-de Sitter black holes in gauged N = 8
  supergravity}},
  \href{http://dx.doi.org/10.1016/S0550-3213(99)00299-0}{\emph{Nucl. Phys.}
  {\bf B554} (1999) 237--253}, [\href{http://arxiv.org/abs/hep-th/9901149}{{\tt
  hep-th/9901149}}].

\bibitem{Cvetic:1999xp}
M.~Cvetic, M.~J. Duff, P.~Hoxha, J.~T. Liu, H.~Lu, J.~X. Lu et~al.,
  \emph{{Embedding AdS black holes in ten-dimensions and eleven-dimensions}},
  \href{http://dx.doi.org/10.1016/S0550-3213(99)00419-8}{\emph{Nucl. Phys.}
  {\bf B558} (1999) 96--126}, [\href{http://arxiv.org/abs/hep-th/9903214}{{\tt
  hep-th/9903214}}].

\bibitem{Andrianopoli:1996cm}
L.~Andrianopoli, M.~Bertolini, A.~Ceresole, R.~D'Auria, S.~Ferrara, P.~Fre
  et~al., \emph{{N=2 supergravity and N=2 superYang-Mills theory on general
  scalar manifolds: Symplectic covariance, gaugings and the momentum map}},
  \href{http://dx.doi.org/10.1016/S0393-0440(97)00002-8}{\emph{J. Geom. Phys.}
  {\bf 23} (1997) 111--189}, [\href{http://arxiv.org/abs/hep-th/9605032}{{\tt
  hep-th/9605032}}].

\bibitem{Cacciatori:2009iz}
S.~L. Cacciatori and D.~Klemm, \emph{{Supersymmetric AdS(4) black holes and
  attractors}}, \href{http://dx.doi.org/10.1007/JHEP01(2010)085}{\emph{JHEP}
  {\bf 01} (2010) 085}, [\href{http://arxiv.org/abs/0911.4926}{{\tt
  0911.4926}}].

\bibitem{Klemm:2012yg}
D.~Klemm and O.~Vaughan, \emph{{Nonextremal black holes in gauged supergravity
  and the real formulation of special geometry}},
  \href{http://dx.doi.org/10.1007/JHEP01(2013)053}{\emph{JHEP} {\bf 01} (2013)
  053}, [\href{http://arxiv.org/abs/1207.2679}{{\tt 1207.2679}}].

\bibitem{Toldo:2012ec}
C.~Toldo and S.~Vandoren, \emph{{Static nonextremal AdS4 black hole
  solutions}}, \href{http://dx.doi.org/10.1007/JHEP09(2012)048}{\emph{JHEP}
  {\bf 09} (2012) 048}, [\href{http://arxiv.org/abs/1207.3014}{{\tt
  1207.3014}}].

\bibitem{Benini1}
F.~Benini, K.~Hristov and A.~Zaffaroni, \emph{{Black hole microstates in
  AdS$_4$ from supersymmetric localization}},
  \href{http://arxiv.org/abs/1511.04085}{{\tt 1511.04085}}.

\bibitem{Hristov1}
K.~Hristov, A.~Tomasiello and A.~Zaffaroni, \emph{{Supersymmetry on
  Three-dimensional Lorentzian Curved Spaces and Black Hole Holography}},
  \href{http://dx.doi.org/10.1007/JHEP05(2013)057}{\emph{JHEP} {\bf 05} (2013)
  057}, [\href{http://arxiv.org/abs/1302.5228}{{\tt 1302.5228}}].

\bibitem{Benini2}
F.~Benini and A.~Zaffaroni, \emph{{A topologically twisted index for
  three-dimensional supersymmetric theories}},
  \href{http://dx.doi.org/10.1007/JHEP07(2015)127}{\emph{JHEP} {\bf 07} (2015)
  127}, [\href{http://arxiv.org/abs/1504.03698}{{\tt 1504.03698}}].

\bibitem{Hosseini1}
S.~M. Hosseini and A.~Zaffaroni, \emph{{Large $N$ matrix models for 3d ${\cal
  N}=2$ theories: twisted index, free energy and black holes}},
  \href{http://arxiv.org/abs/1604.03122}{{\tt 1604.03122}}.

\bibitem{Jokela:2013qya}
N.~Jokela, A.~V. Ramallo and D.~Zoakos, \emph{{Magnetic catalysis in flavored
  ABJM}}, \href{http://dx.doi.org/10.1007/JHEP02(2014)021}{\emph{JHEP} {\bf 02}
  (2014) 021}, [\href{http://arxiv.org/abs/1311.6265}{{\tt 1311.6265}}].

\bibitem{Bea:2016fcj}
Y.~Bea, N.~Jokela and A.~V. Ramallo, \emph{{Quantum phase transitions with
  dynamical flavors}},  \href{http://arxiv.org/abs/1604.03665}{{\tt
  1604.03665}}.

\bibitem{Goldstein:2009cv}
K.~Goldstein, S.~Kachru, S.~Prakash and S.~P. Trivedi, \emph{{Holography of
  Charged Dilaton Black Holes}},
  \href{http://dx.doi.org/10.1007/JHEP08(2010)078}{\emph{JHEP} {\bf 08} (2010)
  078}, [\href{http://arxiv.org/abs/0911.3586}{{\tt 0911.3586}}].

\bibitem{Goldstein:2010aw}
K.~Goldstein, N.~Iizuka, S.~Kachru, S.~Prakash, S.~P. Trivedi and A.~Westphal,
  \emph{{Holography of Dyonic Dilaton Black Branes}},
  \href{http://dx.doi.org/10.1007/JHEP10(2010)027}{\emph{JHEP} {\bf 10} (2010)
  027}, [\href{http://arxiv.org/abs/1007.2490}{{\tt 1007.2490}}].

\bibitem{Amoretti:2016cad}
A.~Amoretti, M.~Baggioli, N.~Magnoli and D.~Musso, \emph{{Chasing the Cuprates
  with Dilatonic Dyons}},  \href{http://arxiv.org/abs/1603.03029}{{\tt
  1603.03029}}.

\bibitem{Papadimitriou:2007sj}
I.~Papadimitriou, \emph{{Multi-Trace Deformations in AdS/CFT: Exploring the
  Vacuum Structure of the Deformed CFT}},
  \href{http://dx.doi.org/10.1088/1126-6708/2007/05/075}{\emph{JHEP} {\bf 05}
  (2007) 075}, [\href{http://arxiv.org/abs/hep-th/0703152}{{\tt
  hep-th/0703152}}].

\bibitem{Gnecchi:2014cqa}
A.~Gnecchi and C.~Toldo, \emph{{First order flow for non-extremal AdS black
  holes and mass from holographic renormalization}},
  \href{http://dx.doi.org/10.1007/JHEP10(2014)075}{\emph{JHEP} {\bf 10} (2014)
  075}, [\href{http://arxiv.org/abs/1406.0666}{{\tt 1406.0666}}].

\bibitem{Witten:2001ua}
E.~Witten, \emph{{Multitrace operators, boundary conditions, and AdS / CFT
  correspondence}},  \href{http://arxiv.org/abs/hep-th/0112258}{{\tt
  hep-th/0112258}}.

\bibitem{Berkooz:2002ug}
M.~Berkooz, A.~Sever and A.~Shomer, \emph{{'Double trace' deformations,
  boundary conditions and space-time singularities}},
  \href{http://dx.doi.org/10.1088/1126-6708/2002/05/034}{\emph{JHEP} {\bf 05}
  (2002) 034}, [\href{http://arxiv.org/abs/hep-th/0112264}{{\tt
  hep-th/0112264}}].

\bibitem{Hertog:2004ns}
T.~Hertog and G.~T. Horowitz, \emph{{Designer gravity and field theory
  effective potentials}},
  \href{http://dx.doi.org/10.1103/PhysRevLett.94.221301}{\emph{Phys. Rev.
  Lett.} {\bf 94} (2005) 221301},
  [\href{http://arxiv.org/abs/hep-th/0412169}{{\tt hep-th/0412169}}].

\bibitem{Freedman:2013ryh}
D.~Z. Freedman and S.~S. Pufu, \emph{{The holography of $F$-maximization}},
  \href{http://dx.doi.org/10.1007/JHEP03(2014)135}{\emph{JHEP} {\bf 03} (2014)
  135}, [\href{http://arxiv.org/abs/1302.7310}{{\tt 1302.7310}}].

\bibitem{Bobev:2011rv}
N.~Bobev, A.~Kundu, K.~Pilch and N.~P. Warner, \emph{{Minimal Holographic
  Superconductors from Maximal Supergravity}},
  \href{http://dx.doi.org/10.1007/JHEP03(2012)064}{\emph{JHEP} {\bf 03} (2012)
  064}, [\href{http://arxiv.org/abs/1110.3454}{{\tt 1110.3454}}].

\bibitem{Gursoy:2007er}
U.~Gursoy, E.~Kiritsis and F.~Nitti, \emph{{Exploring improved holographic
  theories for QCD: Part II}},
  \href{http://dx.doi.org/10.1088/1126-6708/2008/02/019}{\emph{JHEP} {\bf 02}
  (2008) 019}, [\href{http://arxiv.org/abs/0707.1349}{{\tt 0707.1349}}].

\bibitem{Gursoy:2007cb}
U.~Gursoy and E.~Kiritsis, \emph{{Exploring improved holographic theories for
  QCD: Part I}},
  \href{http://dx.doi.org/10.1088/1126-6708/2008/02/032}{\emph{JHEP} {\bf 02}
  (2008) 032}, [\href{http://arxiv.org/abs/0707.1324}{{\tt 0707.1324}}].

\bibitem{Carroll:1997ar}
  S.~M.~Carroll,
  \emph{Lecture notes on general relativity},
  gr-qc/9712019.

\bibitem{Miransky:2015ava}
V.~A. Miransky and I.~A. Shovkovy, \emph{{Quantum field theory in a magnetic
  field: From quantum chromodynamics to graphene and Dirac semimetals}},
  \href{http://dx.doi.org/10.1016/j.physrep.2015.02.003}{\emph{Phys. Rept.}
  {\bf 576} (2015) 1--209}, [\href{http://arxiv.org/abs/1503.00732}{{\tt
  1503.00732}}].

\bibitem{Gusynin:1994re}
V.~P. Gusynin, V.~A. Miransky and I.~A. Shovkovy, \emph{{Catalysis of dynamical
  flavor symmetry breaking by a magnetic field in (2+1)-dimensions}},
  \href{http://dx.doi.org/10.1103/PhysRevLett.73.3499}{\emph{Phys. Rev. Lett.}
  {\bf 73} (1994) 3499--3502}, [\href{http://arxiv.org/abs/hep-ph/9405262}{{\tt
  hep-ph/9405262}}].

\bibitem{Gusynin:1995nb}
V.~P. Gusynin, V.~A. Miransky and I.~A. Shovkovy, \emph{{Dimensional reduction
  and catalysis of dynamical symmetry breaking by a magnetic field}},
  \href{http://dx.doi.org/10.1016/0550-3213(96)00021-1}{\emph{Nucl. Phys.} {\bf
  B462} (1996) 249--290}, [\href{http://arxiv.org/abs/hep-ph/9509320}{{\tt
  hep-ph/9509320}}].

\bibitem{Pisarski:1984dj}
R.~D. Pisarski, \emph{{Chiral Symmetry Breaking in Three-Dimensional
  Electrodynamics}},
  \href{http://dx.doi.org/10.1103/PhysRevD.29.2423}{\emph{Phys. Rev.} {\bf D29}
  (1984) 2423}.

\bibitem{Appelquist:1986fd}
T.~W. Appelquist, M.~J. Bowick, D.~Karabali and L.~C.~R. Wijewardhana,
  \emph{{Spontaneous Chiral Symmetry Breaking in Three-Dimensional QED}},
  \href{http://dx.doi.org/10.1103/PhysRevD.33.3704}{\emph{Phys. Rev.} {\bf D33}
  (1986) 3704}.

\bibitem{Klebanov:1999tb}
I.~R. Klebanov and E.~Witten, \emph{{AdS / CFT correspondence and symmetry
  breaking}},
  \href{http://dx.doi.org/10.1016/S0550-3213(99)00387-9}{\emph{Nucl. Phys.}
  {\bf B556} (1999) 89--114}, [\href{http://arxiv.org/abs/hep-th/9905104}{{\tt
  hep-th/9905104}}].

\bibitem{Berti}
 E.~Berti, V.~Cardoso and A.~O.~Starinets,
  \emph{{Quasinormal modes of black holes and black branes}},
  Class.\ Quant.\ Grav.\  {\bf 26} (2009) 163001
  doi:10.1088/0264-9381/26/16/163001
  [arXiv:0905.2975 [gr-qc]].

\bibitem{Starinets:2002br}
A.~O. Starinets, \emph{{Quasinormal modes of near extremal black branes}},
  \href{http://dx.doi.org/10.1103/PhysRevD.66.124013}{\emph{Phys. Rev.} {\bf
  D66} (2002) 124013}, [\href{http://arxiv.org/abs/hep-th/0207133}{{\tt
  hep-th/0207133}}].

\bibitem{Edalati:2010hk}
M.~Edalati, J.~I. Jottar and R.~G. Leigh, \emph{{Shear Modes, Criticality and
  Extremal Black Holes}},
  \href{http://dx.doi.org/10.1007/JHEP04(2010)075}{\emph{JHEP} {\bf 04} (2010)
  075}, [\href{http://arxiv.org/abs/1001.0779}{{\tt 1001.0779}}].

\bibitem{Witten:1998zw}
E.~Witten, \emph{{Anti-de Sitter space, thermal phase transition, and
  confinement in gauge theories}}, {\emph{Adv. Theor. Math. Phys.} {\bf 2}
  (1998) 505--532}, [\href{http://arxiv.org/abs/hep-th/9803131}{{\tt
  hep-th/9803131}}].

\bibitem{Maldacena:1998im}
J.~M. Maldacena, \emph{{Wilson loops in large N field theories}},
  \href{http://dx.doi.org/10.1103/PhysRevLett.80.4859}{\emph{Phys. Rev. Lett.}
  {\bf 80} (1998) 4859--4862}, [\href{http://arxiv.org/abs/hep-th/9803002}{{\tt
  hep-th/9803002}}].

\bibitem{Rey:1998ik}
S.-J. Rey and J.-T. Yee, \emph{{Macroscopic strings as heavy quarks in large N
  gauge theory and anti-de Sitter supergravity}},
  \href{http://dx.doi.org/10.1007/s100520100799}{\emph{Eur. Phys. J.} {\bf C22}
  (2001) 379--394}, [\href{http://arxiv.org/abs/hep-th/9803001}{{\tt
  hep-th/9803001}}].

\bibitem{Ryu:2006bv}
S.~Ryu and T.~Takayanagi, \emph{{Holographic derivation of entanglement entropy
  from AdS/CFT}},
  \href{http://dx.doi.org/10.1103/PhysRevLett.96.181602}{\emph{Phys. Rev.
  Lett.} {\bf 96} (2006) 181602},
  [\href{http://arxiv.org/abs/hep-th/0603001}{{\tt hep-th/0603001}}].

\bibitem{Klebanov:2007ws}
I.~R. Klebanov, D.~Kutasov and A.~Murugan, \emph{{Entanglement as a probe of
  confinement}},
  \href{http://dx.doi.org/10.1016/j.nuclphysb.2007.12.017}{\emph{Nucl. Phys.}
  {\bf B796} (2008) 274--293}, [\href{http://arxiv.org/abs/0709.2140}{{\tt
  0709.2140}}].

\bibitem{Bali:2012zg}
G.~S. Bali, F.~Bruckmann, G.~Endrodi, Z.~Fodor, S.~D. Katz and A.~Schafer,
  \emph{{QCD quark condensate in external magnetic fields}},
  \href{http://dx.doi.org/10.1103/PhysRevD.86.071502}{\emph{Phys. Rev.} {\bf
  D86} (2012) 071502}, [\href{http://arxiv.org/abs/1206.4205}{{\tt
  1206.4205}}].

\bibitem{Preis:2010cq}
F.~Preis, A.~Rebhan and A.~Schmitt, \emph{{Inverse magnetic catalysis in dense
  holographic matter}},
  \href{http://dx.doi.org/10.1007/JHEP03(2011)033}{\emph{JHEP} {\bf 03} (2011)
  033}, [\href{http://arxiv.org/abs/1012.4785}{{\tt 1012.4785}}].

\bibitem{Preis:2012fh}
F.~Preis, A.~Rebhan and A.~Schmitt, \emph{{Inverse magnetic catalysis in field
  theory and gauge-gravity duality}},
  \href{http://dx.doi.org/10.1007/978-3-642-37305-3_3}{\emph{Lect. Notes Phys.}
  {\bf 871} (2013) 51--86}, [\href{http://arxiv.org/abs/1208.0536}{{\tt
  1208.0536}}].

\bibitem{Mamo:2015dea}
K.~A. Mamo, \emph{{Inverse magnetic catalysis in holographic models of QCD}},
  \href{http://dx.doi.org/10.1007/JHEP05(2015)121}{\emph{JHEP} {\bf 05} (2015)
  121}, [\href{http://arxiv.org/abs/1501.03262}{{\tt 1501.03262}}].

\bibitem{Ferreira:2014kpa}
M.~Ferreira, P.~Costa, O.~Louren\c{c}o, T.~Frederico and C.~Provid\^{e}ncia,
  \emph{{Inverse magnetic catalysis in the (2+1)-flavor Nambu-Jona-Lasinio and
  Polyakov-Nambu-Jona-Lasinio models}},
  \href{http://dx.doi.org/10.1103/PhysRevD.89.116011}{\emph{Phys. Rev.} {\bf
  D89} (2014) 116011}, [\href{http://arxiv.org/abs/1404.5577}{{\tt
  1404.5577}}].

\bibitem{Donos:2011pn}
A.~Donos, J.~P. Gauntlett and C.~Pantelidou, \emph{{Magnetic and Electric AdS
  Solutions in String- and M-Theory}},
  \href{http://dx.doi.org/10.1088/0264-9381/29/19/194006}{\emph{Class. Quant.
  Grav.} {\bf 29} (2012) 194006}, [\href{http://arxiv.org/abs/1112.4195}{{\tt
  1112.4195}}].

\bibitem{Donos:2011qt}
A.~Donos, J.~P. Gauntlett and C.~Pantelidou, \emph{{Spatially modulated
  instabilities of magnetic black branes}},
  \href{http://dx.doi.org/10.1007/JHEP01(2012)061}{\emph{JHEP} {\bf 01} (2012)
  061}, [\href{http://arxiv.org/abs/1109.0471}{{\tt 1109.0471}}].

\bibitem{Almuhairi:2011ws}
A.~Almuhairi and J.~Polchinski, \emph{{Magnetic AdS x R$^2$: Supersymmetry and
  stability}},  \href{http://arxiv.org/abs/1108.1213}{{\tt 1108.1213}}.

\bibitem{Hartnoll:2009sz}
S.~A. Hartnoll, \emph{{Lectures on holographic methods for condensed matter
  physics}},
  \href{http://dx.doi.org/10.1088/0264-9381/26/22/224002}{\emph{Class. Quant.
  Grav.} {\bf 26} (2009) 224002}, [\href{http://arxiv.org/abs/0903.3246}{{\tt
  0903.3246}}].

\bibitem{McGreevy:2009xe}
J.~McGreevy, \emph{{Holographic duality with a view toward many-body physics}},
  \href{http://dx.doi.org/10.1155/2010/723105}{\emph{Adv. High Energy Phys.}
  {\bf 2010} (2010) 723105}, [\href{http://arxiv.org/abs/0909.0518}{{\tt
  0909.0518}}].

\bibitem{Herzog:2009xv}
C.~P. Herzog, \emph{{Lectures on Holographic Superfluidity and
  Superconductivity}},
  \href{http://dx.doi.org/10.1088/1751-8113/42/34/343001}{\emph{J. Phys.} {\bf
  A42} (2009) 343001}, [\href{http://arxiv.org/abs/0904.1975}{{\tt
  0904.1975}}].

\bibitem{Sachdev:2010ch}
S.~Sachdev, \emph{{Condensed Matter and AdS/CFT}},
  \href{http://arxiv.org/abs/1002.2947}{{\tt 1002.2947}}.

\bibitem{Khveshchenko:2001zza}
D.~V. Khveshchenko, \emph{{Magnetic field-induced insulating behavior in highly
  oriented pyrolitic graphite}},
  \href{http://dx.doi.org/10.1103/PhysRevLett.87.206401}{\emph{Phys. Rev.
  Lett.} {\bf 87} (2001) 206401},
  [\href{http://arxiv.org/abs/cond-mat/0106261}{{\tt cond-mat/0106261}}].

\bibitem{Kopelevich}
Y.~e.~a. Kopelevich, \emph{{}}, {\emph{J. Low Temp. Phys} {\bf 119} (2000)
  691}.

\bibitem{Dall'Agata:2010gj}
G.~Dall'Agata and A.~Gnecchi, \emph{{Flow equations and attractors for black
  holes in N = 2 U(1) gauged supergravity}},
  \href{http://dx.doi.org/10.1007/JHEP03(2011)037}{\emph{JHEP} {\bf 03} (2011)
  037}, [\href{http://arxiv.org/abs/1012.3756}{{\tt 1012.3756}}].

\bibitem{Hristov:2010ri}
K.~Hristov and S.~Vandoren, \emph{{Static supersymmetric black holes in $AdS_4$
  with spherical symmetry}},
  \href{http://dx.doi.org/10.1007/JHEP04(2011)047}{\emph{JHEP} {\bf 04} (2011)
  047}, [\href{http://arxiv.org/abs/1012.4314}{{\tt 1012.4314}}].

\bibitem{Dall'Agata:2012bb}
G.~Dall'Agata, G.~Inverso and M.~Trigiante, \emph{{Evidence for a family of
  SO(8) gauged supergravity theories}},
  \href{http://dx.doi.org/10.1103/PhysRevLett.109.201301}{\emph{Phys. Rev.
  Lett.} {\bf 109} (2012) 201301}, [\href{http://arxiv.org/abs/1209.0760}{{\tt
  1209.0760}}].

\bibitem{Johnson:1999qt}
C.~V. Johnson, A.~W. Peet and J.~Polchinski, \emph{{Gauge theory and the
  excision of repulsone singularities}},
  \href{http://dx.doi.org/10.1103/PhysRevD.61.086001}{\emph{Phys. Rev.} {\bf
  D61} (2000) 086001}, [\href{http://arxiv.org/abs/hep-th/9911161}{{\tt
  hep-th/9911161}}].

\bibitem{deBoer:1999tgo}
J.~de~Boer, E.~P. Verlinde and H.~L. Verlinde, \emph{{On the holographic
  renormalization group}},
  \href{http://dx.doi.org/10.1088/1126-6708/2000/08/003}{\emph{JHEP} {\bf 08}
  (2000) 003}, [\href{http://arxiv.org/abs/hep-th/9912012}{{\tt
  hep-th/9912012}}].

\bibitem{deHaro:2000vlm}
S.~de~Haro, S.~N. Solodukhin and K.~Skenderis, \emph{{Holographic
  reconstruction of space-time and renormalization in the AdS / CFT
  correspondence}},
  \href{http://dx.doi.org/10.1007/s002200100381}{\emph{Commun. Math. Phys.}
  {\bf 217} (2001) 595--622}, [\href{http://arxiv.org/abs/hep-th/0002230}{{\tt
  hep-th/0002230}}].

\bibitem{Bianchi:2001kw}
M.~Bianchi, D.~Z. Freedman and K.~Skenderis, \emph{{Holographic
  renormalization}},
  \href{http://dx.doi.org/10.1016/S0550-3213(02)00179-7}{\emph{Nucl. Phys.}
  {\bf B631} (2002) 159--194}, [\href{http://arxiv.org/abs/hep-th/0112119}{{\tt
  hep-th/0112119}}].

\bibitem{Papadimitriou:2005ii}
I.~Papadimitriou and K.~Skenderis, \emph{{Thermodynamics of asymptotically
  locally AdS spacetimes}},
  \href{http://dx.doi.org/10.1088/1126-6708/2005/08/004}{\emph{JHEP} {\bf 08}
  (2005) 004}, [\href{http://arxiv.org/abs/hep-th/0505190}{{\tt
  hep-th/0505190}}].

\bibitem{Batrachenko:2004fd}
A.~Batrachenko, J.~T. Liu, R.~McNees, W.~A. Sabra and W.~Y. Wen, \emph{{Black
  hole mass and Hamilton-Jacobi counterterms}},
  \href{http://dx.doi.org/10.1088/1126-6708/2005/05/034}{\emph{JHEP} {\bf 05}
  (2005) 034}, [\href{http://arxiv.org/abs/hep-th/0408205}{{\tt
  hep-th/0408205}}].

\bibitem{Hawking:1995ap}
S.~W. Hawking and S.~F. Ross, \emph{{Duality between electric and magnetic
  black holes}}, \href{http://dx.doi.org/10.1103/PhysRevD.52.5865}{\emph{Phys.
  Rev.} {\bf D52} (1995) 5865--5876},
  [\href{http://arxiv.org/abs/hep-th/9504019}{{\tt hep-th/9504019}}].

\bibitem{Ashtekar:1984zz}
A.~Ashtekar and A.~Magnon, \emph{{Asymptotically anti-de Sitter space-times}},
  \href{http://dx.doi.org/10.1088/0264-9381/1/4/002}{\emph{Class. Quant. Grav.}
  {\bf 1} (1984) L39--L44}.

\bibitem{Ashtekar:1999jx}
A.~Ashtekar and S.~Das, \emph{{Asymptotically Anti-de Sitter space-times:
  Conserved quantities}},
  \href{http://dx.doi.org/10.1088/0264-9381/17/2/101}{\emph{Class. Quant.
  Grav.} {\bf 17} (2000) L17--L30},
  [\href{http://arxiv.org/abs/hep-th/9911230}{{\tt hep-th/9911230}}].

\bibitem{Kinar:1998vq}
Y.~Kinar, E.~Schreiber and J.~Sonnenschein, \emph{{Q anti-Q potential from
  strings in curved space-time: Classical results}},
  \href{http://dx.doi.org/10.1016/S0550-3213(99)00652-5}{\emph{Nucl. Phys.}
  {\bf B566} (2000) 103--125}, [\href{http://arxiv.org/abs/hep-th/9811192}{{\tt
  hep-th/9811192}}].

\end{thebibliography}
\end{document}